\documentclass[review]{elsarticle}
\usepackage[fleqn]{amsmath}
\usepackage{graphicx}
\graphicspath{ {images/} }
\usepackage[margin=1in]{geometry}
\usepackage{tikz}
\usepackage{tikz-3dplot}
\usepackage{textcomp}
\usepackage{epstopdf}
\usepackage{indentfirst}
\usepackage[section]{placeins}
\usepackage[labelfont=bf,justification=justified,singlelinecheck=false]{caption}
\usepackage{subcaption}
\usepackage{comment}
\usepackage{siunitx}
\usepackage{tabularx}
\usepackage{natbib}
\biboptions{numbers,sort&compress}
\usepackage{color}
 \allowbreak
 \allowdisplaybreaks
\usepackage[inline, shortlabels]{enumitem}
\setlist{itemjoin ={,\enspace},itemjoin* = {, and\enspace}}
\usepackage{soul}
\usepackage{multirow}

\usepackage{floatrow,array}%
\usepackage{hyperref}
\usepackage{lineno}
\newcommand*\patchAmsMathEnvironmentForLineno[1]{%
	\expandafter\let\csname old#1\expandafter\endcsname\csname #1\endcsname
	\expandafter\let\csname oldend#1\expandafter\endcsname\csname end#1\endcsname
	\renewenvironment{#1}%
	{\linenomath\csname old#1\endcsname}%
	{\csname oldend#1\endcsname\endlinenomath}}%
\newcommand*\patchBothAmsMathEnvironmentsForLineno[1]{%
	\patchAmsMathEnvironmentForLineno{#1}%
	\patchAmsMathEnvironmentForLineno{#1*}}%
\AtBeginDocument{%
	\patchBothAmsMathEnvironmentsForLineno{equation}%
	\patchBothAmsMathEnvironmentsForLineno{align}%
	\patchBothAmsMathEnvironmentsForLineno{flalign}%
	\patchBothAmsMathEnvironmentsForLineno{alignat}%
	\patchBothAmsMathEnvironmentsForLineno{gather}%
	\patchBothAmsMathEnvironmentsForLineno{multline}%
}
\modulolinenumbers[5]
\begin{document}

\title{Invariant compact finite difference schemes}

\author{E.~Ozbenli\corref{cor1}}
\ead{ozbenli@ou.edu}

\author{P.~Vedula\corref{}}

\address{School of Aerospace and Mechanical Engineering, University of Oklahoma,
865 Asp Ave., Norman, OK73019, USA}

 \begin{abstract}\label{sec:abs}
 
 In this paper, we propose a method, that is based on equivariant moving frames, for development of high order accurate invariant compact finite difference schemes that preserve Lie symmetries of underlying partial differential equations. 
 In this method, variable transformations that are obtained from the extended symmetry groups of PDEs are used to transform independent and dependent variables, and derivative terms of compact finite difference schemes (constructed for these PDEs) such that the resulting schemes are invariant under the chosen symmetry groups. 
 The unknown symmetry parameters that arise from the application of these transformations are determined through selection of convenient moving frames.
 In some cases, owing to selection of convenient moving frames, numerical representation of invariant (or symmetry preserving) compact numerical schemes, that are developed through the proposed method, are found to be notably simpler than that of standard, noninvariant compact numerical schemes.
 Further, the order of accuracy of these invariant compact schemes can be arbitrarily set to a desired order by considering suitable compact finite difference algorithms. 
 Application of the proposed method is demonstrated through construction of invariant compact finite difference schemes for some common linear and nonlinear PDEs (including the linear advection-diffusion equation in 1D/2D, the inviscid and viscous Burgers' equations in 1D).
 Results obtained from our numerical simulations indicate that invariant compact finite difference schemes not only inherit selected symmetry properties of underlying PDEs, but are also comparably more accurate than the standard, non-invariant base numerical schemes considered here.

\end{abstract}

\maketitle


  \section{Introduction}\label{sec:intro}
  
  Compact finite differencing based on Pad\'{e} approximants is a commonly used high order numerical method that is well-documented in the literature \cite{hirsh1975,lele1992,ciment1975,mahesh1998,dai2000,sherer2005,shukla2007,rizzetta2008,cui2009,shah2010}.
  An important objective of this method is to achieve high order accuracy with a relatively small number of stencil points by relating a weighted sum of functions (or dependent variables) to a weighted sum of derivatives, evaluated at grid points.
  Hence, numerical solutions based on compact schemes are found to have good, spectral-like resolutions, solutions that exponentially converge with increasing resolution, especially in the case of short waves \cite{hirsh1975}.
  In this regard, Hirsh \cite{hirsh1975} presented a detailed application of compact finite differencing which included development and application of fourth order accurate compact schemes to three test problems, namely viscous Burgers' equation, Howarth's retarded boundary layer flow, and the incompressible driven cavity problem.
  The author also provided a brief discussion on how to treat boundary conditions when developing compact finite difference schemes, which could be problematic in some cases.
  In another work, Lele \cite{lele1992} extended the earlier works on compact finite differencing by presenting finite difference schemes that provide a better representation of shorter length scales for use on problems with a range of spatial scales. 
  In addition, the author provided a detailed discussion on how to obtain compact finite difference schemes of various orders (up to tenth order) and treat the relevant boundary conditions.
  In a more recent work, Shukla \textit{et al}. \cite{shukla2007} presented a family of high order compact schemes that are built on non-uniform grids with spatial orders of accuracy ranging from 4th to 20th. These compact schemes are constructed such that they maintain high-order accuracy not only in the interior of a domain, but also at its boundaries. The authors demonstrated the application of these compact schemes to the linear wave equation and two-dimensional incompressible Navier-Stokes equations, and verified the achievement of high order accuracy for these problems. They further showed (via comparisons with benchmark solutions for the two-dimensional driven cavity flow, thermal convection in a square box, and flow past an impulsively started cylinder) that these high order compact schemes are stable and produce highly accurate results on stretched grids with more points clustered at boundaries. 
  
  Although compact finite differencing is an efficient method for construction of high order accurate numerical schemes, these schemes often ignore geometric properties of underlying differential equations as the focus is usually on the accuracy when developing these schemes. 
  Schemes that preserve certain geometric properties (such as energy, momentum, symplecticity, Hamiltonian and Poisson structures of equations) are usually considered as geometric integrators.
  %
  It is well-documented in literature that geometric integrators, which account for certain geometric properties of underlying differential equations, are likely to perform better than schemes standard schemes that ignore such properties \cite{zhong1988,yoshida1990,mclachlan1993,calvo1995,islas2001,hairer2006,webb2014,gagarina2016}. 

  Lie symmetry groups of differential equations are also geometric properties that could be preserved in numerical schemes. 
  Numerous researchers have proposed methods for construction of numerical schemes that preserve symmetry groups of underlying differential equations \cite{yanenko1976,dorodnitsyn1993,bakirova1997,Dorodnitsyn2000,dorodnitsyn2003,dorodnitsyn2010,dorodnitsyn2017,budd1997,hydon2000,budd2001,levi2005,bourlioux2006,Hoarau2006,levi2014,fels1998,fels1999,kim2006,kim2007,kim2008,chhay2010,chhay2011,bihlo2017,ozbenli2017a,ozbenli2017b,ozbenli2018consideration}.
  Most of these works can be categorized into two major groups. In the first group \cite{dorodnitsyn1993,bakirova1997,Dorodnitsyn2000,dorodnitsyn2003,dorodnitsyn2010}, invariants of difference equations are determined through Lie's infinitesimal approach, and then, a set of these invariants are used to construct invariant schemes that converge to the original differential equations in the continuous limit. 
  In the other group \cite{fels1998,fels1999,kim2006,kim2007,kim2008,chhay2010,chhay2011,bihlo2017,ozbenli2017a,ozbenli2017b,ozbenli2018consideration}, point transformations based on symmetry groups of differential equations are applied to some base (non-invariant) numerical schemes, and the unknown symmetry parameters of these transformations are determined through moving frames that are based on Cartan's method of normalization \cite{cartan5methode}. 
  
  In this paper, we propose a mathematical approach for construction of high order accurate compact finite difference schemes that retain Lie symmetry groups of underlying differential equations.
  In the proposed method that is based on equivariant moving frames, extended symmetry groups of partial differential equations are used to obtain point transformations not only for independent and dependent variables of differential equations, but also for their derivative terms (which is a novel aspect of this paper that was not considered in earlier works \cite{kim2006,kim2007,ozbenli2017b}).
  Once point transformations for derivatives of differential equations are determined, then these transformations are applied to some (non-invariant) base compact finite difference schemes (of a desired order of accuracy) to obtain final invariant (or symmetry preserving) forms of these schemes. 
  Here, we note that the unknown symmetry parameters that appear in these point transformations are determined by choosing convenient moving frames for which numerical representations of base schemes simplify notably, and their accuracy improves. 
  The proposed method is applied to some commonly used linear and nonlinear problems, and for all the test problems, the resulting invariant schemes are found to perform significantly better than selected non-invariant base compact schemes in terms of numerical accuracy, verifying the potential advantages of symmetry preservation.

  We demonstrate the implementation of the proposed method by considering fourth order accurate invariant compact finite difference schemes for one- and two-dimensional linear advection-diffusion equations and Burgers' equations (i.e., inviscid, viscous).
  For numerical simplicity, we use forward differencing to discretize temporal derivatives, and fourth order compact schemes based on central differencing to discretize spatial derivatives.
  Note that the proposed construction of invariant schemes can also be extended to arbitrarily high order temporal and spatial discretizations.  
  Results obtained from the proposed invariant compact schemes developed for these test problems suggest that symmetry preservation can lead to significant improvements in numerical accuracy, besides storing important geometric information (regarding the underlying differential equations) in associated numerical schemes.
  
  This paper is organized as follows. In Section \ref{sec:Symm}, the formulation for the fourth order accurate compact schemes along with a detailed discussion on Lie symmetry analysis and invariantization of compact schemes are provided. The step by step development of invariant compact schemes for some linear and nonlinear problems are noted in Section \ref{sec:invnumsch}. Performance of the constructed invariant compact schemes, along with a detailed discussion of the results obtained from these schemes are presented in Section \ref{results}. And finally, the concluding remarks and a brief summary of the work are given in Section \ref{sec:conc}.

  \section{Mathematical formulation}\label{sec:Symm}
 
 In this section, the procedure (that is based on equivariant moving frames) for construction of invariant compact schemes is presented in detail. Brief discussions on Lie symmetry analysis and compact schemes are also included.

 \subsection{Construction of compact schemes}\label{sec:lie}
 
 Compact finite difference methods are widely used for high order computations, and in some cases are favored over standard finite difference methods, due to their ability to achieve high order accuracy over smaller stencils. 
 For instance, while a standard central difference approximation of the first derivative of a function on a three-point stencil is second order accurate, an approximation based on a compact scheme (that is also derived through central differencing) of the same derivative could be of higher orders. 
 The implementation of compact schemes is rather simple.
 To illustrate construction of compact schemes through an example, let us develop fourth order accurate compact finite difference schemes for the first and second derivatives of a function $U$. 
 Consider the following Taylor series expansion of the function U at grid points $(i \pm 1)$:
 \begin{align}
 	U^{i\pm 1}=&U^{i}\pm h U_x^{i}+\frac{h^2}{2}U_{xx}^{i}\pm \frac{h^3}{6}U_{xxx}^{i}+\frac{h^4}{24}U_{(\rm IV)}^{i}\pm O(h^5) \label{uip1}
 \end{align}
 where $h$ is the discrete spatial step and the symbol $(\cdot)_{x}$ denotes derivative with respect to variable $x$. Similarly, the first and second derivative of $U$ can be expanded in a Taylor series as  
 \begin{align}
 	U_{x}^{i\pm 1}=&U_{x}^{i}\pm h U_{xx}^{i}+\frac{h^2}{2}U_{xxx}^{i}\pm \frac{h^3}{6}U_{(\rm IV)}^{i}+\frac{h^4}{24}U_{(\rm V)}^{i}\pm O(h^5) \label{uxip1}\\
 	U_{xx}^{i\pm 1}=&U_{xx}^{i}\pm h U_{xxx}^{i}+\frac{h^2}{2}U_{(\rm IV)}^{i}\pm \frac{h^3}{6}U_{(\rm V)}^{i}+\frac{h^4}{24}U_{(\rm VI)}^{i}\pm O(h^5) \label{uxxip1}.
 \end{align}
 In order to eliminate the higher order derivatives (i.e., $U_{xx},U_{xxx},U_{(\rm IV)}$, and $U_{(\rm V)}$) and obtain an implicit relationship between the first derivative $U_{x}$ and the function $U$ at nodes $i\pm 1$, one can multiply Eq.~\eqref{uip1} with constant $a$ at point $i+1$, and with constant $b$ at point $i-1$, and multiply Eq.~\eqref{uxip1} with quantity $c\times h$ at point $i+1$, and with quantity $d\times h$ at point $i-1$, and sum up these resulting quantities to obtain the following equation:
 \begin{align} \label{uximp}
 	a U^{i+1}+b U^{i-1}+c h U_{x}^{i+1}+d h U_{x}^{i-1}=& (a+b) U^{i}+ (a-b+c+d) h U_x^{i}+(a+b+2c-2d) \frac{h^2}{2}U_{xx}^{i}\nonumber \\
 	&+(a-b+3c+3d)\frac{h^3}{6}U_{xxx}^{i}+(a+b+4c-4d)\frac{h^4}{24}U_{(\rm IV)}^{i}\nonumber\\
 	&+(c+d)\frac{h^5}{24}U_{(\rm IV)}^{i}+O(h^5). 
 \end{align}
 The arbitrary constants $a$, $b$, $c$, and $d$ can be obtained via elimination of high order derivatives as  $ \{ a,b,c,d \}=\{ \frac{3}{4},-\frac{3}{4},-\frac{1}{4},-\frac{1}{4} \}$. Hence, the final form of Eq.~\eqref{uximp} is
 %
 %
 %
 %
 \begin{align} \label{uxcomp}
 	\frac{1}{6}U_{x}^{i+1}+\frac{2}{3}U_{x}^{i}+\frac{1}{6}U_{x}^{i-1}=\frac{U^{i+1}-U^{i-1}}{2h}+O(h^4)~
 \end{align}
 which relates the function U to its first derivative and provides a fourth order accurate implicit approximation for the first derivative of $U$. 
 Through similar algebraic manipulations, one can obtain the following fourth order accurate implicit approximation for the second derivative of the function U as well
 \begin{align} \label{uxxcomp}
 	\frac{1}{12}U_{xx}^{i+1}+\frac{5}{6}U_{xx}^{i}+\frac{1}{12}U_{xx}^{i-1}=\frac{U^{i+1}-2U^{i}+U^{i-1}}{h^2}+O(h^4).
 \end{align}
 Both Eqs.~\eqref{uxcomp}--\eqref{uxxcomp} yield tridiagonal matrices that can easily be solved to accurately approximate the first and second derivatives of $U$ at all grid points.
 More information on compact schemes along with compact algorithms for derivatives with higher orders of accuracy, and a discussion on the treatment of boundary conditions in this approach can be found in the literature \cite{hirsh1975,lele1992}.

 \subsection{Lie symmetry analysis}
 
 A differential equation is said to possess a symmetry property if one can transform every variable in the equation according to some transformations, such that the resulting output reads exactly the same as the original differential equation in new (transformed) variables. 
 Further, a Lie point symmetry group is an algebraic structure that consists of a set of objects, which correspond to continuous symmetries (or coordinate transformations) that map a system to itself with a binary operation that satisfies the following group axioms:
 \begin {enumerate*} [1) ]%
 \item closure \item existence of identity element \item existence of inverse element \item associativity.
 \end {enumerate*} 
 The procedure for determination of Lie point symmetries of equations is straightforward and well-documented in the literature \cite{ibragimov1995,cantwell2002,ovsiannikov2014,vu2012,oberlack1999}.
 
 In this context, consider a surface $L(\mathbf{x},\mathbf{u},\mathbf{p})=0$ to be a partial differential equation, and let the following be a one-parameter ($k$th-extended) Lie group $G$:
 \begin{align} \label{1parsymmgrp}
 	\tilde{x}^j\,&=\,\tilde{x}^j(\mathbf{x},\mathbf{u},s)\nonumber\\
 	\tilde{u}^i\,&=\,\tilde{u}^i(\mathbf{x},\mathbf{u},s)\nonumber\\
 	\tilde{u}^i_{j_1}\,&=\,\tilde{u}^i_{j_1}(\mathbf{x},\mathbf{u},\mathbf{u}_1,s)\\
 	&\vdots\nonumber\\
 	\tilde{u}^i_{j_1j_2\ldots j_k}\,&=\,\tilde{u}^i_{j_1j_2\ldots j_k}(\mathbf{x},\mathbf{u},\mathbf{p},s)\nonumber
 \end{align}
 where the arbitrary constant $s$ is the symmetry (or group) parameter, and $\mathbf{p}=(\mathbf{u}_1,\mathbf{u}_2, \ldots, \mathbf{u}_k)$. Here, the vectors $\mathbf{x}=(x^1,x^2, \ldots, x^m)$ and $\mathbf{u}=(u^1,u^2, \ldots, u^n)$ denote the independent and dependent variables, respectively, and $\mathbf{u}_k$ represents the vector of all possible $k$th order derivatives of $\mathbf{u}$ with respect to the independent variables. Also, the operator ${(\cdot)}_{j_1j_2\cdots j_k}$ represents the partial derivative $\frac{\partial^k (\cdot)}{\partial x_{j_1} \partial x_{j_2}\cdots \partial x_{j_k} }$. The smooth transformation functions ($\tilde{x}^j$, $\tilde{u}^i,\ldots \tilde{u}^i_{j_1j_2\ldots, j_k}$) given in group $G$ can be further expanded in a Taylor series about the point $s=0$ to determine the infinitesimal form of the one-parameter Lie group $G$ as 
 \begin{align} \label{1parsymmgrpinf}
 	\tilde{x}^j&=x^j+s\,[\xi^j(\mathbf{x},\mathbf{u})]+O(s^2)~, \quad \quad  \xi^j\equiv \left [\frac{\partial \tilde{x}^j}{\partial s}\right ]_{s=0} \nonumber\\
 	\tilde{u}^i&=u^i+s\,[\eta^i(\mathbf{x},\mathbf{u})]+O(s^2)~, \quad \quad \, \eta^i\equiv \left [\frac{\partial \tilde{u}^i}{\partial s}\right ]_{s=0}\nonumber\\
 	\tilde{u}^i_{j_1}&={u}^i_{j_1}+s\,[\eta^i_{[j_1]} (\mathbf{x},\mathbf{u},\mathbf{u}_1)]+O(s^2)~, \quad \quad  \eta^i_{[j_1]} \equiv \left [\frac{\partial \tilde{u}^i_{j_1}}{\partial s}\right ]_{s=0}\\
 	&\vdots\nonumber\\
 	\tilde{u}^i_{j_1j_2\ldots j_k}&={u}^i_{j_1j_2\ldots j_k}+s\,[\eta^i_{[j_1\ldots j_k]} (\mathbf{x},\mathbf{u},\mathbf{p})]+O(s^2)~, \quad  \eta^i_{[j_1\ldots j_k]} \equiv \left [\frac{\partial \tilde{u}^i_{j_1j_2\ldots j_k}}{\partial s}\right ]_{s=0}~.\nonumber
 \end{align}
 where $\xi^j$ and $\eta^i$ are known as the coordinate functions (or the group infinitesimals), which define the transformation of the coordinate variables under the action of the group $G$. Similarly, $\eta^i_{[j_1\cdots j_k]}$ is the $k$th-extended group infinitesimal that defines how the $k$th derivative is transformed under the action of $G$, and is given by the following relation:
 \begin{align}
 	\eta^i_{[j_1\ldots j_k]} = D_{j_k}\eta^i_{[j_1\ldots j_{k-1}]}-u^i_{j_1\ldots j_{k-1} r} D_{j_k} \xi^r
 \end{align}
 where $D_{j_k}$ is the total derivative operator \cite{cantwell2002}.
 
 The surface $L(\mathbf{x},\mathbf{u},\mathbf{p})=0$ is said to be invariant under the action of the group $G$ if the equation reads the same in new variables as represented below:
 \begin{equation} \label{invariance}
 	L(\mathbf{x},\;\mathbf{u},\;\mathbf{p})=0 \quad \Longleftrightarrow \quad L(\tilde{\mathbf{x}},\;\tilde{\mathbf{u}},\;\tilde{\mathbf{p}})=0~.
 \end{equation}
 In order to determine the Lie point symmetry group $G$ that will leave the surface $L(\mathbf{x},\mathbf{u},\mathbf{p})=0$ invariant (or unchanged), the following invariance condition is applied 
 \begin{equation} \label{invcond}
 	\mathbf{X}_{[k]}\,\,L(\mathbf{x},\;\mathbf{u},\;\mathbf{p})=0~, \quad \quad (\text{mod}\; L = 0)
 \end{equation}
 where $\mathbf{X}_{[k]}$ is the $k$-extended group operator that is of the form 
 \begin{equation} \label{kthgrpopr}
 	\mathbf{X}_{[k]}=\xi^j \frac{\partial}{\partial x^j}+\eta^i \frac{\partial}{\partial u^i}+ \cdots + \eta^i_{[j_1\ldots j_k]} \frac{\partial}{\partial {u}^i_{j_1j_2\ldots j_k}}~.
 \end{equation}
 Solution of the invariance condition given in Eq.~\eqref{invcond} through determination of the coordinate functions yields the Lie point symmetry group $G$ associated with the surface $L(\mathbf{x},\mathbf{u},\mathbf{p})=0$. 
 A more detailed discussion on Lie symmetry analysis, particularly regarding how to solve the invariance condition, can be found in the reference \cite{cantwell2002}.

 \subsection{Invariantization of compact schemes}
 
 In this work, a compact finite difference scheme (corresponding to a surface $L(\mathbf{z})=0$) is considered as an invariant compact scheme if its form remains unchanged under the action of a point symmetry group $G$ associated with the surface $L(\mathbf{z})=0$. In this context, let $\tilde N_c(\mathbf{z})=0$ be an invariant compact finite difference scheme, and $\tilde \phi_c(\mathbf{z})=0$ be a stencil equation for the surface $L(\mathbf{z})=0$ where $\mathbf{z}=(\mathbf{x},\mathbf{u},\mathbf{p})$ is the vector of the independent/dependent variables and derivatives, respectively. The compact scheme $\tilde N_c(\mathbf{z})=0$ and the stencil equation $\tilde \phi_c(\mathbf{z})=0$ are said to be invariant under the action of the group element $g$ (where $g \in G$) if the following condition is satisfied \cite{fels1999,kim2006,kim2007}:
 \begin{equation} \label{invcompschms}
 	\begin{split}
 		\tilde N_c(\rho(\mathbf{z}) \cdotp \mathbf{z})&=0 \quad \Longleftrightarrow \quad  \tilde N_c(\mathbf{z})=0 \\
 		\tilde \phi_c(\rho(\mathbf{z}) \cdotp \mathbf{z})&=0 \quad \Longleftrightarrow \; \quad  \tilde \phi_c(\mathbf{z})=0
 	\end{split}
 \end{equation}
 where $\rho(\mathbf{z})$ represents right moving frames defined on a manifold $M$ such that it is a topological map ($\rho: M\rightarrow G$) that satisfies the following condition:
 \begin{align}
 	\rho(g\,\cdotp \mathbf{z})\,=\,\rho(\mathbf{z})\,g^{-1}\nonumber
 \end{align}
 for $\forall\,g\,\in\,G$. For any given non-invariant compact finite difference scheme $N_c(\mathbf{z})=0$ (constructed for a surface $L(\mathbf{z})=0$), an invariant form of this scheme $\tilde N_c(\mathbf{z})=0$ can be obtained by simply transforming every coordinate variable and derivative of the base (non-invariant) compact scheme according to the symmetry group $G$ as $\tilde N_c(\mathbf{z})=N_c(g\cdot\mathbf{z})$ for all $g \in G$. The unknown group parameters (that appear when the action of a particular group element $g$ on the coordinate variables and derivatives is evaluated) can be determined via Cartan's method of normalization. A more detailed discussion on Cartan's method of normalization and equivariant moving frames can be found in the literature \cite{cartan5methode,fels1998,fels1999,kim2006}.

 \section{Construction of invariant numerical schemes}\label{sec:invnumsch}
In this section, the invariantization of compact finite difference schemes is illustrated through examples. In particular, fourth order accurate invariant compact schemes are constructed for some linear and nonlinear problems.


\subsection{Inviscid Burgers' equation} \label{subsec:IBE1D}

As our first test problem, we consider the inviscid Burgers' equation (IBE), which is a model that describes nonlinear wave propagation, and is of the form
\begin{align} \label{ibe1dform}
	u_t\:+\:u\:u_x=0.
\end{align}
A non-invariant compact scheme can be constructed for the IBE using the compact algorithms developed for the spatial first, Eq.~\eqref{uxcomp}, and second, Eq.~\eqref{uxxcomp}, derivatives. As for the time derivative, for simplicity, a classical first order forward differencing technique can be considered. The order of accuracy can be improved from first to second order via truncation error analysis or defect correction. Hence the final form of the compact scheme develop for the inviscid Burgers' equation can be found as
\begin{align} \label{ibecompsch}
	N_c(\mathbf{z})=\frac{u^{(i,n+1)}-u^{(i,n)}}{\tau}+u u_x + d_c = 0~. 
\end{align}
Here, $d_c$ represents the defect correction terms (obtained from truncation error analysis) that are added to the scheme to improve accuracy and is given by
\begin{align} \label{defcor}
	d_c=-\frac{\tau}{2}(u^2\:u_{xx}+2\:u\:u^2_x)\,\,+\,\,\,O(\tau^2,h^4)~
\end{align}
where $\tau$ and $h$ denote the discrete time and space steps, respectively.

Further, the symmetry group $G$ associated with the inviscid Burgers' equation can be found (via Lie symmetry analysis) as
\begin{align}\label{ibesymgro}
	X_1\,&=\,t^2\:\frac{\partial}{\partial \:t}\,+\,x\:t\:\frac{\partial}{\partial \:x}\,+\,(x-t\,u)\:\frac{\partial}{\partial \:u}\,\nonumber\\
	X_2\,&=\,t\,x\:\frac{\partial}{\partial \:t}\,+\,x^2\:\frac{\partial}{\partial \:x}\,+u\,(x-t\,u)\:\frac{\partial}{\partial \:u}\,\nonumber\\
	X_3\,&=2\,t\,\frac{\partial}{\partial \:t}\,+x\:\frac{\partial}{\partial \:x}\,-u\:\frac{\partial}{\partial \:u}\,\nonumber\\
	X_4\,&=\,x\,\frac{\partial}{\partial \:t}\,-\,u^2\:\frac{\partial}{\partial \:u}\,\\
	X_5\,&=\,t\,\frac{\partial}{\partial \:x}\,+\:\frac{\partial}{\partial \:u}\,\nonumber\\
	X_6\,&=\,\frac{\partial}{\partial \:t}\,\nonumber\\
	X_7\,&=\,\frac{\partial}{\partial \:x}\,\nonumber
\end{align}
where $X_{r=1,\ldots,7}$, is the group operator that corresponds to that particular subgroup. The point transformations, $ \tilde{\mathbf{z}}= (\tilde t, \tilde x, \tilde u, \tilde{\mathbf{p}} )$, associated with a particular subgroup can be found using Lie series expansion as follows
\begin{align}\label{lieser}
	\tilde{z}_i\,=\,e^{(s_j X_j)}\;z_i\,=\,z_i+s_j\:(X_j\:z_i)+\frac{s_j^2}{2!}\:X_j(X_j\:z_i)+\frac{s_j^3}{3!}\:X_j(X_j(X_j\:z_i))+\cdots~.
\end{align}
Here, we note that in order to find the extended point transformations $ \tilde{\mathbf{p}}= (\tilde u_{\tilde x},\tilde u_{\tilde x \tilde x})$, one should extend the group operators given in Eq.~\eqref{ibesymgro} such that it accounts for all the derivative terms before these groups are used in the Lie series given in Eq.~\eqref{lieser}. Alternatively, one can use the chain rule to find the extended point transformations. For instance, the transformation expression for the spatial first derivative can be found using 
\begin{align}
	\frac{\partial \tilde{u}}{\partial \tilde{x}}=\frac{\partial \tilde{u}}{\partial x}\frac{\partial {x}}{\partial \tilde{x}}+\frac{\partial \tilde{u}}{\partial t}\frac{\partial {t}}{\partial \tilde{x}} \nonumber \nonumber
\end{align}
once the point transformations for the independent and dependent variables are found.
Similarly, point transformations associated with a multiple number of subgroups can be obtained by substituting each subgroup into Eq.~\eqref{lieser} in an arbitrary order. Although it is possible to consider the full Lie algebra and obtain global transformations for the coordinate variables and derivatives, it is sometimes practical to choose only certain subgroups as the form of the point transformations obtained from the full Lie algebra could be cumbersome and not practical for preservation in associated compact finite difference schemes \cite{ozbenli2017b}. Hence, for this particular problem, we only choose the subgroups $X_1$, $X_3$, $X_6$, and $X_7$ for preservation in the associated (non-invariant) compact scheme given in Eq.~\eqref{ibecompsch}. The global transformations obtained from these particular subgroups are found via Eq.~\eqref{lieser} as
\begin{align} \label{ibepointtra}
	\tilde t &= e^{2s_3}\frac{(t+s_6)}{\lambda}\nonumber\\
	\tilde x &= e^{s_3}\frac{x+s_7}{\lambda}\nonumber\\
	\tilde u &= e^{-s_3}(\lambda u+s_1(x+s_7))\\
	\tilde u_{\tilde x} &= e^{-2s_3}(\lambda^2 u_x+s_1\lambda)\nonumber\\
	\tilde u_{\tilde x \tilde x} &= e^{-3s_3}\lambda^3 u_{xx}\nonumber
\end{align}
where $\lambda = 1-s_1 (t+s_6)$. The compact scheme constructed for the inviscid Burgers' equation, Eq.~\eqref{ibecompsch}, can be invariantized by transforming every coordinate variable and derivative according to the above transformations 
\begin{align} \label{ibecompschtrans}
	\tilde N_c(\mathbf{z})=N_c(g\cdot\mathbf{z})=\frac{\tilde u^{(i,n+1)}-\tilde u^{(i,n)}}{\tilde{\tau}}+\tilde u \tilde u_{\tilde x} -\frac{\tilde{\tau}}{2}(\tilde u^2\:\tilde u_{\tilde x \tilde x}+2\:\tilde u\:\tilde u^2_{\tilde{x}})\,\,+\,\,\,O(\tilde{\tau}^2,\tilde h^4) = 0~. 
\end{align}
Based on the point transformations given in Eq.~\eqref{ibepointtra}, it appears that the symmetry parameter $s_3$ does not appear in the transformed scheme given in Eq.~\eqref{ibecompschtrans}. All the other symmetry parameters can be determined through Cartan's method of normalization. First, we consider convenient normalization conditions that lead to simple stencils. For instance, normalization conditions $\tilde t^{(i,n)}=0$ and $\tilde x^{(i,n)}=0$, among infinite possibilities, yield a simple stencil where the symmetry parameters $s_6$ and $s_7$ are $-t^{(i,n)}$ and $-x^{(i,n)}$, respectively. Second, we choose normalization conditions that remove terms from the truncation error of compact schemes under consideration and hence lead to a considerable improvement in numerical accuracy, besides simplifying their numerical representations \cite{ozbenli2017b}. In this context, the normalization condition $\tilde u^{(i,n)}_{\tilde x}=0$ can be used to determine the symmetry parameter $s_1$ 
\begin{align} \label{movfrmux}
	\tilde u^{(i,n)}_{\tilde x}=0 \quad  \Rightarrow \quad u^{(i,n)}_x+s_1=0 \quad  \Rightarrow \quad s_1=-u^{(i,n)}_x
\end{align}
as this particular normalization condition removes all the terms that include the spatial first derivative from the compact scheme given in Eq.~\eqref{ibecompschtrans} in the transformed space as shown in the following:
\begin{align} \label{ibeinvcomp}
	\tilde N_c(\mathbf{z})=N_c(g\cdot\mathbf{z})=\frac{\tilde u^{(i,n+1)}-\tilde u^{(i,n)}}{\tilde{\tau}}-\frac{\tilde{\tau}}{2}\tilde u^2\:\tilde u_{\tilde x \tilde x}\,\,+\,\,\,O(\tilde{\tau}^2,\tilde h^4) = 0~. 
\end{align}
The compact scheme given in Eq.~\eqref{ibeinvcomp} is invariant under the symmetry groups $X_1$, $X_3$, $X_6$, and $X_7$ and can also be expressed in original variables as follows
\begin{align} \label{ibeinccomschorg}
	u^{(i,n+1)} =  \frac{1}{\lambda_{n+1}}  \left ( u^{(i,n)} + \frac{\tau^2}{2 \lambda^2_{n+1}}(u^{(i,n)})^2 u_{xx} \right )
\end{align}
where $\lambda_{n+1}=1- s_1 \tau$. 
Note that for most of the test problems considered in this work, we use a time-space orthogonal mesh, $t^{(i+1,n)}-t^{(i,n)}=0$ and $x^{(i,n+1)}-x^{(i,n)}=0$, and hence, for simplicity, we will replace $t^{(i,n)}$ with $t^{n}$, and $x^{(i,n)}$ with $x^{i}$ in the following examples.
Invariance of the compact scheme constructed for the inviscid Burgers' equation, Eq.~\eqref{ibeinccomschorg}, can be verified by transforming every variable in this scheme according to the transformations given in Eq.~\eqref{ibepointtra}, and the resulting transformed scheme should be identical to Eq.~\eqref{ibeinccomschorg}. 

Here we also note that for this particular problem, for simplicity, we considered first order forward differencing for the time derivative and used the method of modified equations to improve the accuracy of the approximation from first to second order. However, one could also use higher order approximations or other discretization techniques (i.e., Runge-Kutta methods) for the time derivative if desired. A particularly interesting case occurs when a TVD-RK2 discretization technique is used for the time derivative in Eq.~\eqref{ibe1dform}. In this case, the final form of the invariant compact scheme constructed using the transformations and moving frames considered for the IBE would be identical to the invariant compact scheme given in Eq.~\eqref{ibeinccomschorg}.


\subsection{Linear advection-diffusion equation in 1D} \label{subsec:ADE1D}

As our second test problem, we choose the one-dimensional linear advection-diffusion equation of the form
\begin{align} \label{ade1dform}
	u_t\:+\:\alpha\:u_x=\nu \:u_{xx}.
\end{align}
which describes the evolution of a quantity $u$ due to linear advection and diffusion processes.
The symbols $\alpha$ and $\nu$ denote the constant characteristic speed and diffusion coefficient, respectively. 
A non-invariant compact numerical scheme can be developed for Eq.~\eqref{ade1dform} as
\begin{align} \label{ade1dcompsch}
	\frac{u^{(i,n+1)}-u^{(i,n)}}{\tau}+\alpha\: u_x  = \nu\:u_{xx}~ 
\end{align}
where forward differencing is considered for the time derivative, and the spatial first and second derivatives are approximated according to Eq.~\eqref{uxcomp} and Eq.~\eqref{uxxcomp}, respectively.
The symmetry group $G$ associated with the one-dimensional advection-diffusion equation is
\begin{align}\label{adesymmetries}
	\begin{split}
		X_1\,&=\,2\:t^2\:\frac{\partial}{\partial \:t}\,+\,2\:x\:t\:\frac{\partial}{\partial \:x}\,-\,u\:(t+\frac{(x-\alpha\:t)^2)}{2\:\nu}\:\frac{\partial}{\partial \:u}\,\\
		X_2\,&=\,4\:t\:\frac{\partial}{\partial \:t}\,+\,2\:(x+\alpha\:t)\:\frac{\partial}{\partial \:x}\,\\ 
		X_3\,&=\,t\:\frac{\partial}{\partial \:x}\,-\,u\:\frac{(x-\alpha\:t)}{2\:\nu}\:\frac{\partial}{\partial \:u}\,\\
		X_4\,&=\,u\:\frac{\partial}{\partial \:u}\,\\
		X_5\,&=\,\frac{\partial}{\partial \:t}\,\\
		X_6\,&=\,\frac{\partial}{\partial \:x}~.
	\end{split}
\end{align}
Considering the subgroups $X_1$, $X_5$, and $X_6$, the following point transformations can be obtained
\begin{align} \label{ade1dvartra}
	\tilde t\:&=\frac{t+s_5}{\lambda} \nonumber\\
	\tilde x\:&=\frac{x+s_6}{\lambda} \nonumber\\ 
	\tilde u\:&= \lambda^{\frac{1}{2}}u\:\exp{\left (-\frac{s_1\:\gamma^2}{2\lambda \nu}\right )} \\
	\tilde u_{\tilde x}\:&=\lambda^{\frac{1}{2}}\nu^{-1} (s_1 \gamma u+\lambda \nu u_x)     \:\exp{\left (-\frac{s_1\:\gamma^2}{2\lambda \nu}\right )} \nonumber\\
	\tilde u_{\tilde x \tilde x}\:&=\lambda^{\frac{1}{2}}\nu^{-2} (s_1^2 \gamma^2 u - s_1 \lambda \nu u- 2s_1 \lambda \gamma \nu u_x  +  \lambda^2 \nu^2 u_{xx})     \:\exp{\left (-\frac{s_1\:\gamma^2}{2\lambda \nu}\right )} \nonumber
\end{align}
where $\lambda=1-2\:s_1\:(t+s_5)$ and $\gamma=x+s_6-\alpha\:(t+s_5)$.
The other subgroups are neglected as their inclusion leads to point transformations of cumbersome structures that are difficult to implement. The normalization conditions $\tilde t^n=0 $ and $\tilde x^i=0$ can be used to determine the symmetry parameters $s_5$ and $s_6$, respectively. The symmetry parameter $s_1$ (corresponding to the projection group $X_1$) can be found by considering the normalization condition
\begin{equation} \label{ade1dprojcond}
	\tilde u^{(i,n)}_{\tilde x \tilde x}=0 \quad \Rightarrow \quad s_1=\frac{\nu}{u^{(i,n)}}\:u^{(i,n)}_{xx}.
\end{equation}
As all the unknown symmetry parameters are defined, the point transformations given in Eq.~\eqref{ade1dvartra} can be implemented to the base compact numerical scheme, Eq.~\eqref{ade1dcompsch}. This implementation appears to reduce the scheme to a form of linear advection equation in the transformed space as 
\begin{align} \label{ade1dtracompsch}
	\frac{\tilde u^{(i,n+1)}-\tilde u^{(i,n)}}{\tilde{\tau}}+\alpha\: \tilde u^{(i,n)}_{\tilde x}  = 0 
\end{align}
where the spatial second derivative is removed from the scheme owing to the normalization condition given in Eq.~\eqref{ade1dprojcond}. 
Hence, the transformed compact scheme given in Eq.~\eqref{ade1dtracompsch} that is constructed for the one-dimensional linear advection-diffusion equation and is invariant under the subgroups $X_1$, $X_5$, and $X_6$ can be expressed in the original discrete variables as follows
\begin{equation}
	u^{(i,n+1)}=\lambda_{n+1}^{-\frac{3}{2}}\:(\lambda_{n+1} u^{(i,n)}-\tau\:\alpha\:u^{(i,n)}_x)\:\exp{\left ( \frac{s_1 \alpha^2 \tau^2}{2\nu \lambda_{n+1}}  \right)}
\end{equation}
where $\lambda_{n+1}=1-2 s_1 \tau$.


\subsection{Viscous Burgers' equation} \label{subsec:VBE1D}

As our third test problem, let us consider the viscous Burgers' equation that is of the form
\begin{align} \label{vbe1dform}
	u_t\:+\:u\:u_x=\nu\:u_{xx}
\end{align}
and develop an invariant compact numerical scheme for this particular PDE. 
Similar to the one-dimensional linear advection-diffusion equation, we consider forward differencing for the time derivative and use Eqs.~\eqref{uxcomp}--\eqref{uxxcomp} for the spatial derivatives to construct the non-invariant base compact scheme for the solution of this PDE as shown in the following
\begin{align} \label{vbecomsch}
	\frac{u^{(i,n+1)}-u^{(i,n)}}{\tau}+u u_x=
	\nu u_{xx}~.
\end{align}
The symmetry group $G$ associated with the viscous Burgers' equation is 
\begin{align}\label{vbesymmetries}
	\begin{split}
		X_1\,&=\,t^2\:\frac{\partial}{\partial \:t}\,+\,x\:t\:\frac{\partial}{\partial \:x}\,+\,(x-t\:u)\:\frac{\partial}{\partial \:u}\,\\
		X_2\,&=\,t\:\frac{\partial}{\partial \:x}\,+\frac{\partial}{\partial \:u}\,\\ 
		X_3\,&=\,2t\:\frac{\partial}{\partial \:t}\,+x\:\frac{\partial}{\partial \:x}\,-u\:\frac{\partial}{\partial \:u}\, \,\\
		X_4\,&=\,\frac{\partial}{\partial \:t}\,\\
		X_5\,&=\,\frac{\partial}{\partial \:x}\,.
	\end{split}
\end{align}
The point transformations that account for the projection group $X_1$, Galilean transformation group $X_2$, scaling group $X_3$, and translation groups $X_4$ and $X_5$ can be found as
\begin{align} \label{vbepointtra}
	\tilde t &= e^{2s_3}\frac{(t+s_4)}{\lambda} \nonumber\\
	\tilde x &= e^{s_3}\frac{x+s_5+s_2 (t+s_4)}{\lambda} \nonumber \\
	\tilde u &= e^{-s_3}(\lambda u+s_1(x+s_5)+s_2)\\
	\tilde u_{\tilde x} &= e^{-2s_3}(\lambda^2 u_x+s_1\lambda) \nonumber \\
	\tilde u_{\tilde x \tilde x} &= e^{-3s_3}\lambda^3 u_{xx} \nonumber
\end{align}
where $\lambda=1-s_1(t+s_4)$. As similar to the inviscid Burgers' equation, the scaling symmetry parameter $s_3$ does not occur when these transformations are implemented to the compact scheme given in Eq.~\eqref{vbecomsch}. The symmetry parameters associated with the translation groups $X_4$ and $X_5$ can be found by considering the same normalization conditions used for the previous problems. The Galilean parameter $s_2$ can be found by using the normalization condition $\tilde u^{(i,n)}=0$. And finally, the projection parameter $s_1$ can be found by choosing a moving frame for which the approximation for the first derivative goes to zero in the transformed space 
\begin{align}\label{vbemovframes1}
	\tilde{u}^{(i,n)}_{\tilde x}=0 \quad  \Rightarrow \quad  s_1=-u^{(i,n)}_{x}~. 
\end{align}
The above normalization condition indicates that all terms in the base (non-invariant) compact scheme, Eq.~\eqref{vbecomsch}, that include the spatial first derivative will be removed from the compact scheme in the transformed space leading to the following reduced form
\begin{align} \label{vbetrasch}
	\tilde u^{(i,n+1)} = \nu\: \tilde{\tau}\: \tilde u_{\tilde x \tilde x}~
\end{align}
where $\tilde{\tau} = \tilde{t}^{(i,n+1)}$.
The transformed compact numerical scheme, Eq.~\eqref{vbetrasch}, that is invariant under all the symmetry groups of the viscous Burgers' equation can also be expressed in original variables as
\begin{align} \label{vbeinvsch}
	u^{(i,n+1)} =  \frac{1}{\lambda_{n+1}} \left ( u^{(i,n)}-s_1 (x^{(i,n+1)}-x^{(i,n)}) + \frac{\tau \nu}{\lambda_{n+1}} u^{(i,n)}_{xx}     \right )              
\end{align}
where $\lambda_{n+1}=1-s_1 \tau$.


\subsection{Advection-diffusion equation in 2D} \label{subsec:ADE2D}

As our last test problem, we choose the two-dimensional linear advection-diffusion equation that is of the form 
\begin{align} \label{ade2dform}
	u_t\:+\:\alpha\,u_x\:+\,\beta\:u_y=\nu\,(u_{xx}+u_{yy})
\end{align}
to demonstrate the applicability of the proposed method to a multidimensional problem. Here $\alpha$ and $\beta$ denote constant characteristic wave speeds along $x$ and $y$ coordinates, respectively.
For this particular PDE, two different compact numerical schemes that are invariant under the same symmetry groups, but are constructed using different moving frames, are developed.
Similar to the previous problems, the base (non-invariant) compact numerical scheme considered for this PDE is also developed considering forward differencing for the temporal derivative and fourth order compact finite difference algorithms, given in Eq.~\eqref{uxcomp}--\eqref{uxxcomp}, for the spatial derivatives as shown in the following:
\begin{align} \label{ade2dcomsch}
	\frac{u^{(i,j,n+1)}-u^{(i,j,n)}}{\tau}+\alpha\, u_x+\beta\, u_y = \nu\, (u_{xx}+u_{yy})~.
\end{align}
Considering the symmetry group associated with the two-dimensional linear advection-diffusion equation,
\begin{align}\label{ade2dsymmetries}
	X_1\,&=\,4\:\nu\:t^2\:\frac{\partial}{\partial \:t}\,
	+\,4\:\nu\:x\:t\:\frac{\partial}{\partial \:x}\,
	+\,4\:\nu\:y\:t\:\frac{\partial}{\partial \:y}\,
	-u\:[(x-\alpha\:t)^2+(y-\beta\:t)^2 +4 \nu t]\:\frac{\partial}{\partial \:u}\,\nonumber\\
	X_2\,&=\,2\:\nu\:t\:\frac{\partial}{\partial \:x}\,+2\:\nu\:t\:\frac{\partial}{\partial \:y}-u\,(x-\alpha\:t+y-\beta\:t)\:\frac{\partial}{\partial \:u}\,\nonumber\\ 
	X_3\,&=\,2\:\nu\:y\:\frac{\partial}{\partial \:x}\,-2\:\nu\:x\:\frac{\partial}{\partial \:y}-u\,(\beta\:x-\alpha\:y)\:\frac{\partial}{\partial \:u}\,\nonumber\\ 
	X_4\,&=\,4\:\nu\:t\:\frac{\partial}{\partial \:t}\,
	+\,2\nu\:x\:\frac{\partial}{\partial \:x}\,
	+\,2\nu\:y \:\frac{\partial}{\partial \:y}\,
	+u[\alpha (x-\alpha\:t)+\beta (y-\beta t)]\frac{\partial}{\partial u}\,\\
	X_5\,&=\,u\:\frac{\partial}{\partial \:u},\nonumber\\
	X_6\,&=\,\frac{\partial}{\partial \:t}\,\nonumber\\
	X_7\,&=\,\frac{\partial}{\partial \:x}\,\nonumber\\
	X_8\,&=\,\frac{\partial}{\partial \:y}\,\nonumber
\end{align}
the following point transformations that are based on the subgroups $X_1$, $X_6$, $X_7$, and $X_8$, are found
\begin{align} \label{ade2dpoitranss}
	\tilde t\:&=\frac{t+s_6}{\lambda}, \quad \quad 
	\tilde x\:=\frac{x+s_7}{\lambda} , \quad \quad 
	\tilde y\:=\frac{y+s_8}{\lambda} \nonumber\\
	\tilde u\:&=\lambda\:u\:\exp{\left [-\frac{ s_1\:(\gamma^2 + \theta^2)}{\lambda} \right ]}\nonumber\\
	\tilde u_{\tilde x}\:&= (2\,\lambda\:\gamma\: s_1\: u\:+\:\lambda^2\: u_x)\: \exp{\left [-\frac{ s_1\:(\gamma^2 + \theta^2)}{\lambda} \right ]} \nonumber\\
	\tilde u_{\tilde y}\:&= (2\,\lambda\:\theta\: s_1\: u\:+\:\lambda^2\: u_y)\: \exp{\left [-\frac{ s_1\:(\gamma^2 + \theta^2)}{\lambda} \right ]}\\
	\tilde u_{\tilde x \tilde x}\:&= (4\,\lambda\:\gamma^2\: s_1^2\: u\:-\:2\,\lambda^2\:s_1\:u\:+\:4\,\lambda^2\:\gamma\:s_1\: u_x\:+\:\lambda^3\:u_{xx})\: \exp{\left [-\frac{ s_1\:(\gamma^2 + \theta^2)}{\lambda} \right ]}\nonumber\\
	\tilde u_{\tilde y \tilde y}\:&= (4\,\lambda\:\theta^2\: s_1^2\: u\:-\:2\,\lambda^2\:s_1\:u\:+\:4\,\lambda^2\:\theta\:s_1\: u_y\:+\:\lambda^3\:u_{yy})\: \exp{\left [-\frac{ s_1\:(\gamma^2 + \theta^2)}{\lambda} \right ]} \nonumber
\end{align}
where
\begin{align*}
	\lambda &= 1-4\:\nu\:s_1\:(t+s_6)\\
	\gamma &=  x+s_7-\alpha\:(t+s_6)\\
	\theta &=  y+s_8-\beta\:(t+s_6)~.
\end{align*}
The base compact scheme given in Eq.~\eqref{ade2dcomsch} can be transformed according to the above transformations as follows 
\begin{align} \label{ade2dcomschtransformed}
	\frac{\tilde u^{(i,j,n+1)}-\tilde u^{(i,j,n)}}{\tilde{\tau}}+\alpha\, \tilde u_{\tilde x}+\beta\, \tilde u_{\tilde y} = \nu \,(\tilde u_{\tilde x \tilde x}+\tilde u_{\tilde y \tilde y})~.
\end{align}
Here we note that, for simplicity, we ignore the Galilean ($X_2$ and $X_3$) and scaling ($X_4$ and $X_5$) groups and do not consider them for determination of the point transformations as their inclusion (besides the other symmetry groups) result in transformations that are laborious to implement. The symmetry parameters $s_6$, $s_7$, and $s_8$ can be determined by considering the normalization conditions $\tilde{t}^n=0$, $\tilde{x}_i=0$, and $\tilde{y}_j=0$, respectively. As for the determination of the symmetry parameter $s_1$, we consider two different normalization conditions to evaluate the effect of these selections on the numerical accuracy of the resulting invariant schemes. We choose 
\begin{align}\label{ade2dnorm1}
	\partial_{\tilde{x} \tilde{x}} \tilde u^{(i,j,n)}=0 \quad  \Rightarrow \quad  s_1=\frac{\partial_{x x} u^{(i,j,n)}}{2\:u^{(i,j,n)}}~  
\end{align}
as the first normalization condition and construct an invariant compact scheme (referred to as SYM-1) as follows
\begin{align} \label{ade2dcomschtrans1}
	\frac{\tilde u^{(i,j,n+1)}-\tilde u^{(i,j,n)}}{\tilde{\tau}}+\alpha \tilde u_{\tilde x}+\beta \tilde u_{\tilde y} = \nu \tilde u_{\tilde y \tilde y}~.
\end{align}
In the second case, we consider the normalization condition
\begin{align}\label{ade2dnorm2}
	\partial_{\tilde{x} \tilde{x}} \tilde u^{(i,j,n)}+\partial_{\tilde{y} \tilde{y}} \tilde u^{(i,j,n)}=0 \quad  \Rightarrow \quad  s_1=\frac{\partial_{x x} u^{(i,j,n)}+\partial_{y y} u^{(i,j,n)}}{4\:u^{(i,j,n)}}~ 
\end{align}
and construct another invariant compact scheme (referred to as SYM-2) as 
\begin{align} \label{ade2dcomschtrans2}
	\frac{\tilde u^{(i,j,n+1)}-\tilde u^{(i,j,n)}}{\tilde{\tau}}+\alpha \tilde u_{\tilde x}+\beta \tilde u_{\tilde y} = 0~.
\end{align}
Here we note that both Eq.~\eqref{ade2dcomschtrans1} and Eq.~\eqref{ade2dcomschtrans2} can also be expressed in the original variables by implementing the transformations given in Eq.~\eqref{ade2dpoitranss}.


 \section{Results} \label{results}
 
 In this section, performance of the proposed invariant compact finite difference schemes developed for the inviscid Burgers' equation, linear advection-diffusion equation (in 1D and 2D), and viscous Burgers' equation is evaluated. Results obtained from these invariant schemes are compared with the standard schemes for numerical accuracy. 
 
 We first evaluate the performance of the invariant compact scheme constructed for the inviscid Burgers' equation, Eq.~\eqref{ibeinccomschorg}, by comparing the results with the analytical solution 
 \begin{align} \label{ibeexact}
 u(t,x)=\frac{1}{\sqrt{2\,\pi\,\sigma^2}} \exp \left [{-\frac{(x-u(t,x)\,t)^2}{2\,\sigma^2}}\right ]
 \end{align}
 over the spatial domain $\Gamma(x)$ where $x \in [-3,3]$. The initial and boundary conditions are noted from the analytical solution. 

 \begin{table}[t]
    \captionsetup{width=.70\textwidth}
   \captionsetup{skip=1pt}
    \centering
    \caption{Root mean square error (RMSE) and $L_\infty$ error associated with numerical solutions for the inviscid Burgers' equation.}\label{table1}
    \renewcommand{\arraystretch}{1.2}   
    \small
    \begin{tabular}{@{\hspace{1.0em}}c@{\hspace{1.5em}}@{\hspace{3.1em}}c@{\hspace{3.1em}}
    @{\hspace{3.0em}}c@{\hspace{3.0em}}@{\hspace{3.1em}}c@{\hspace{1.1em}}}
    \hline
    Error & FTCS & COMP & SYM \\ \hline
    $L_\infty$ & $4.0 \times 10^{-2}$ & $5.8 \times 10^{-3}$ & $5.1 \times 10^{-3}$ \\ 
    RMSE & $9.7 \times 10^{-3}$ & $1.3 \times 10^{-3}$ & $1.1 \times 10^{-3}$ \\ \hline
    \end{tabular}
 \end{table}
 %
 %
   %
  \begin{figure}[t] 
  \centering
  \includegraphics[width=0.48\textwidth]{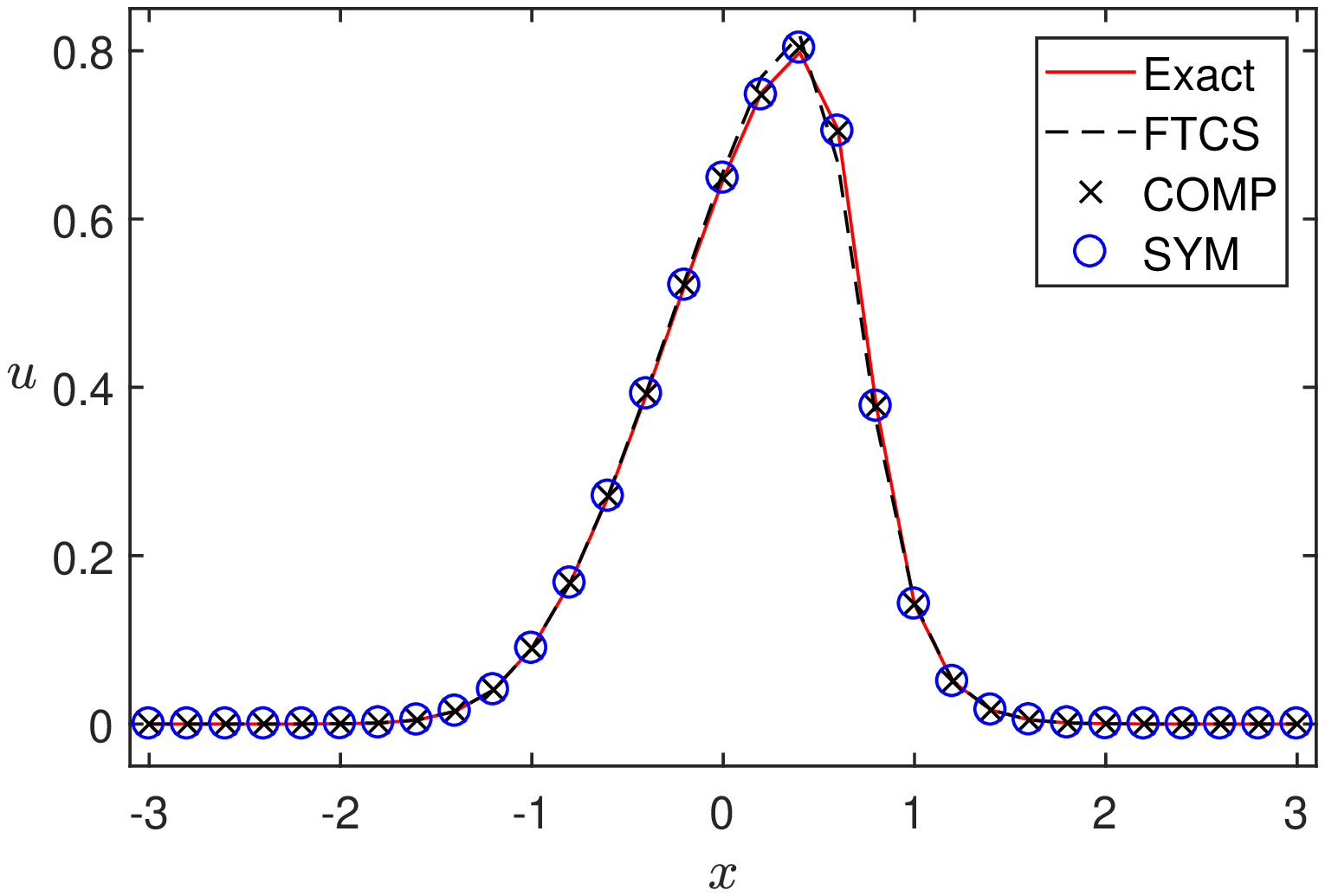} 
  \hfill
  \includegraphics[width=0.48\textwidth]{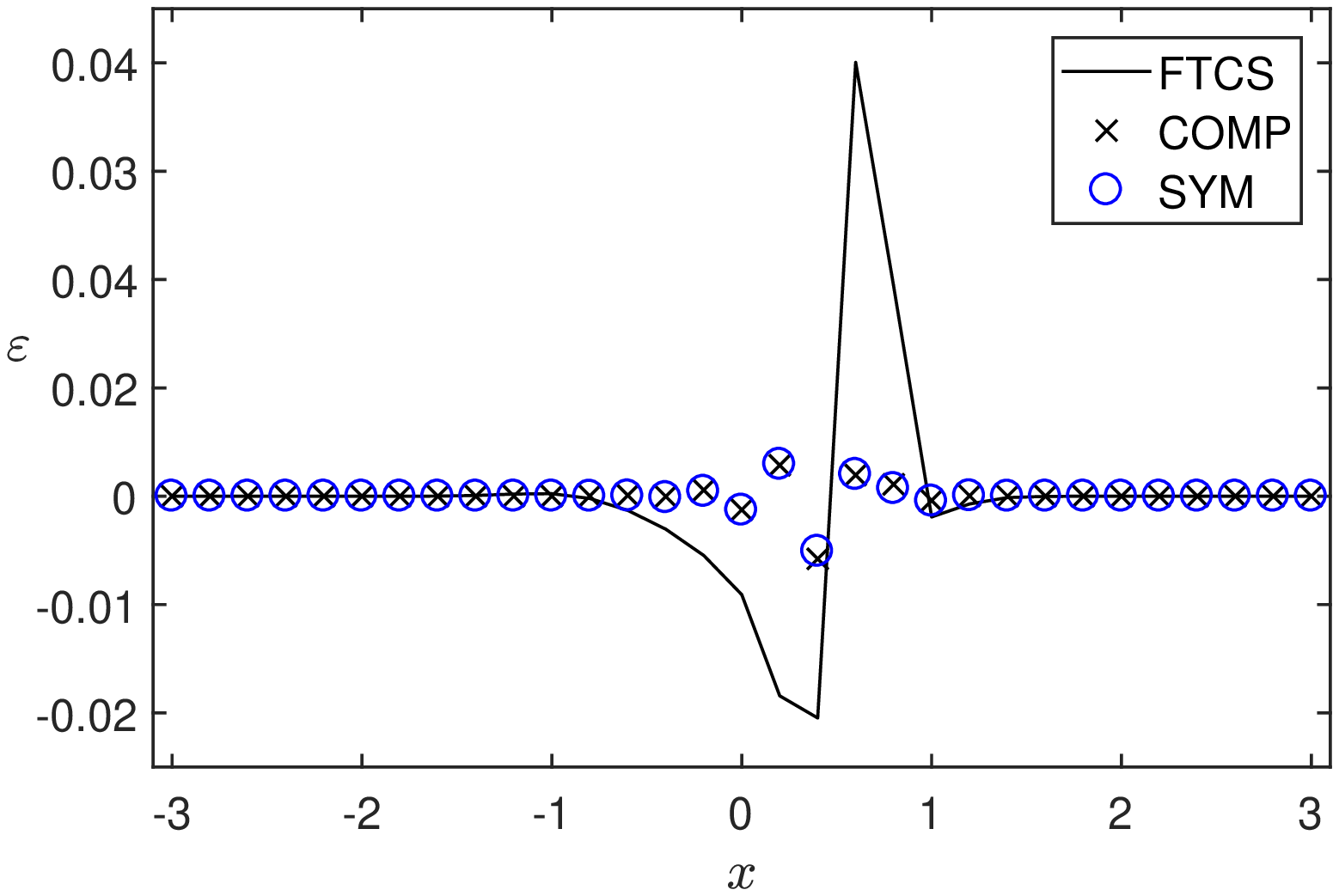} 
  \caption{Inviscid Burgers' equation. Comparison of wave formation profiles, at $t=0.5$, obtained from the analytical solution (Exact), the standard forward in time central in space scheme (FTCS), the standard compact scheme (COMP), and the proposed invariant compact scheme (SYM) is shown in the left figure. Spatial distribution of errors for these numerical schemes is displayed in the right figure. Parameter settings:\,\,$h=0.2$,\,\,$\tau=0.001$, and $\sigma=0.5$~.}
   \label{fig1}
  \end{figure}
 Snapshots of the propagating wave that are obtained from the exact solution, the proposed invariant (compact) scheme (SYM), standard fourth order accurate compact scheme (COMP), and the classical second order accurate forward in time central in space (FTCS) scheme are shown in figure \ref{fig1} (left plot). 
 The associated numerical errors of these schemes, which are estimated as $N_{exact} - N_{numeric}$, are also given in this figure \ref{fig1} (right plot). 
 It appears that the results obtained from the proposed invariant compact scheme (SYM) are significantly more accurate than those obtained from the standard finite difference scheme (FTCS) and are slightly better than those obtained from the standard compact finite difference scheme (COMP).
 Further, the root mean square error (RMSE), estimated as $\sqrt{\sum{(u_{a}-u_{n})^2}/N}$, and $L_\infty$ error, estimated as $max(|u_a-u_n|)$, of these numerical schemes, for this particular run, are given in table \ref{table1}.
 According to the error analysis presented in this table, the $L_\infty$ errors obtained from the FTCS scheme, compact finite difference scheme, and the invariant scheme are $4.0 \times 10^{-2}$, $5.8 \times 10^{-3}$, and $5.1 \times 10^{-3}$, respectively. Similarly, the root mean square errors for these numerical schemes are found as $9.7 \times 10^{-3}$ (FTCS), $1.3 \times 10^{-3}$ (COMP), and $1.1 \times 10^{-3}$ (SYM).
 As expected, the proposed invariant scheme (SYM) which preserves the symmetries of the underlying PDE has significantly less error compared to the standard FTCS scheme and has slightly less error than the standard compact finite difference scheme.

\begin{figure}[htbp!] 
  \centering
  \includegraphics[width=0.75\textwidth]{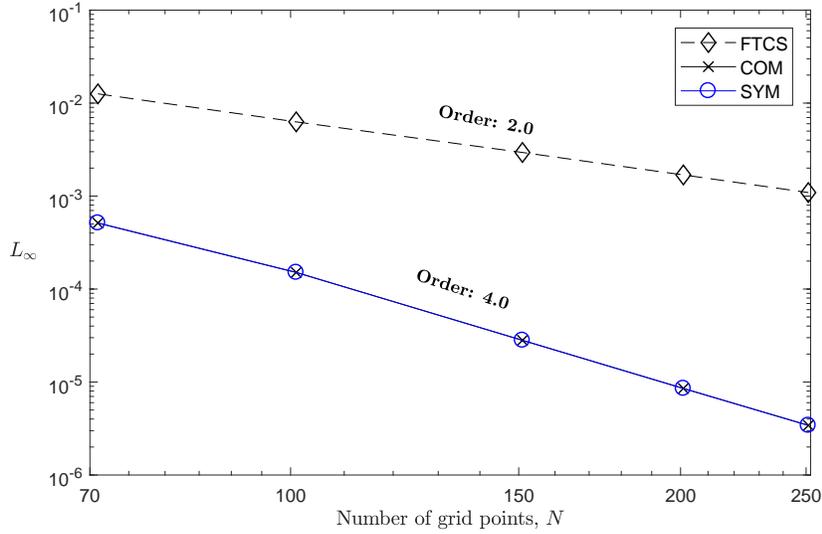} 
  \captionsetup{format=hang}
  \caption{Inviscid Burgers' equation. Comparison of $L_\infty$ errors of numerical schemes as a function of number of grid points.}
   \label{fig2}
  \end{figure}

 The variation of $L_\infty$ errors (obtained from the standard FTCS scheme, standard compact finite difference scheme, and the invariant scheme) with respect to the number of spatial grid points is demonstrated in figure \ref{fig2}. 
 The proposed invariant scheme (SYM) appears to be two orders more accurate than the standard second order FTCS scheme and is at the same order as the standard compact finite difference scheme which is known to be fourth order accurate. Here, we note that a sufficiently small-time step is considered for this simulation as the fourth order compact algorithms (given in Eq.~\eqref{uxcomp} and Eq.~\eqref{uxxcomp}) are only considered for the spatial derivatives.  
  %
 %
 %
  %
  %

 %
 
\begin{figure}[t] 
  	\centering
  	\includegraphics[width=0.48\textwidth]{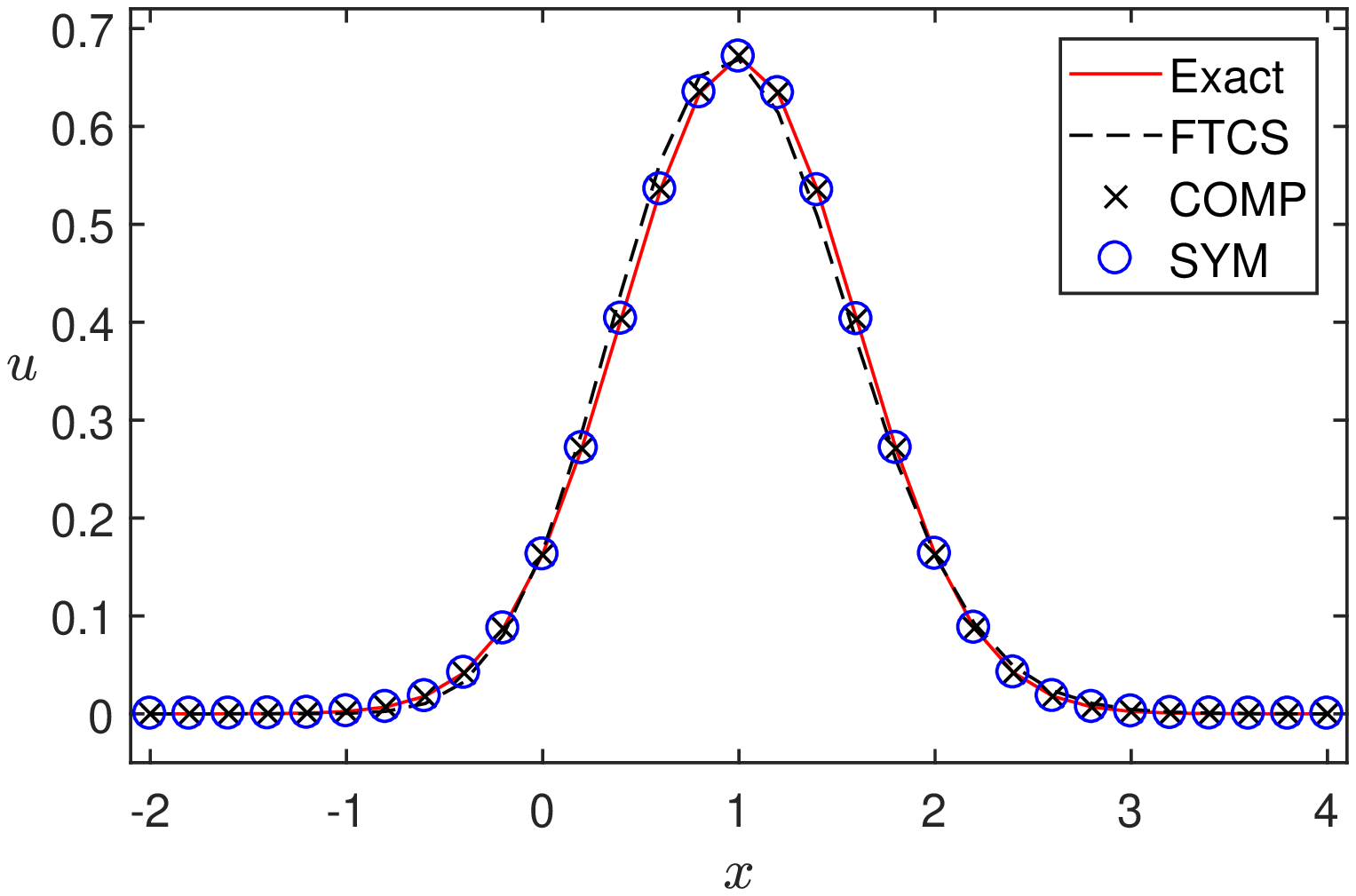} 
  	\hfill
  	\includegraphics[width=0.48\textwidth]{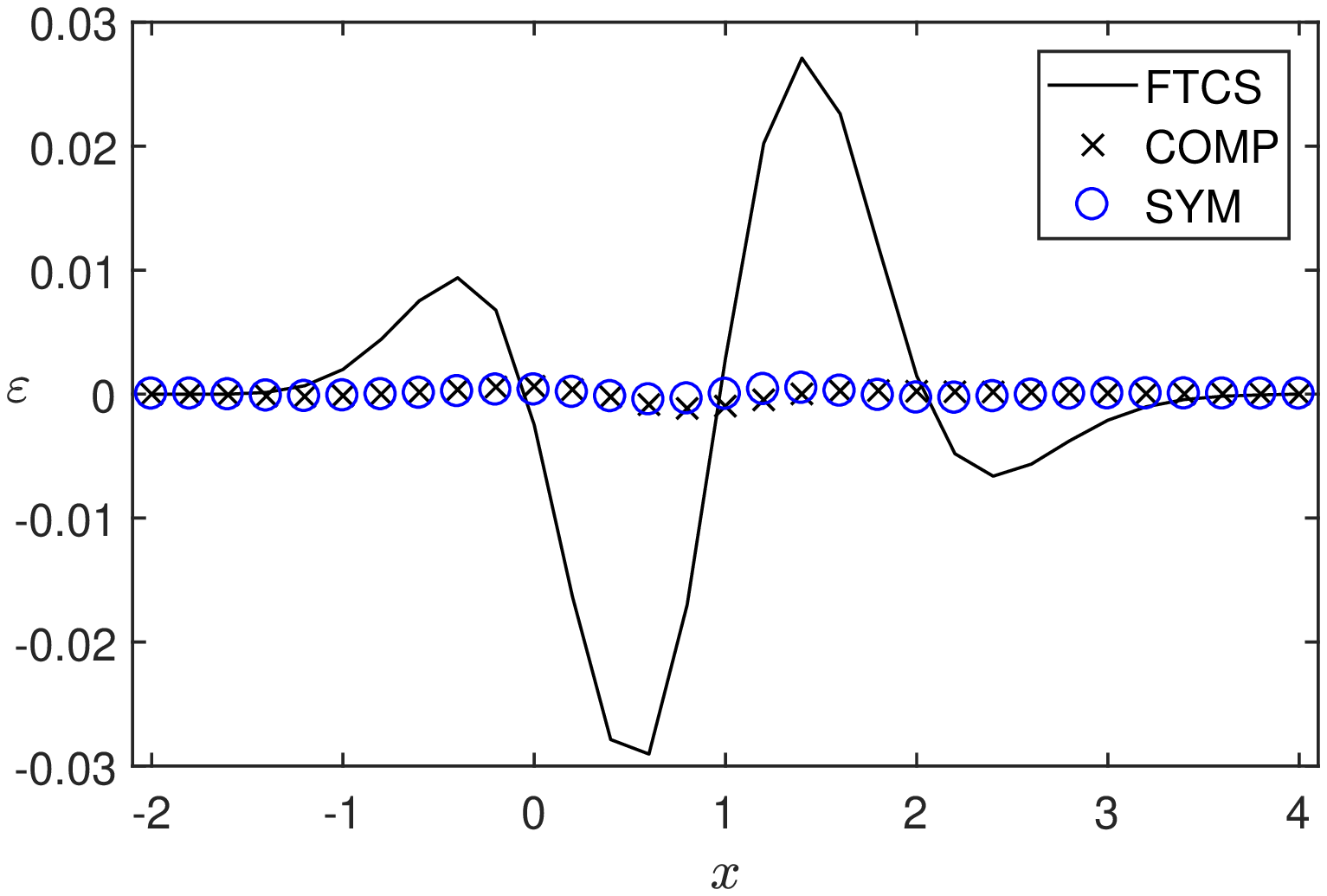} 
  	\captionsetup{format=hang}
  	\caption{Advection-diffusion equation in 1D. Snapshots of wave profiles, at $t=1.0$, obtained from the analytical solution (Exact), the classical forward in time central in space scheme (FTCS), the standard compact scheme (COMP), and the proposed invariant compact scheme (SYM) are displayed in the left figure. Spatial distribution of errors is displayed in the right figure. Parameter settings:\,\,$h=0.2$,\,\,$\tau=0.001$,\,\, $\nu=1/60$.~.}
  	\label{fig3}
  \end{figure}

 Further, we evaluated the performance of the proposed method by developing a fourth order accurate invariant compact finite difference scheme for the one-dimensional linear advection-diffusion equation given in Eq.~\eqref{ade1dform}.
 The following analytical solution
 \begin{align} \label{ade1dexact}
 	u(t,x)=\frac{1}{\sqrt{4\,\pi\,(L^2+\nu\:t)}} \exp \left [{-\frac{(x-\alpha\,t)^2}{4\,\,(L^2+\nu\:t)}}\right ]
 \end{align}
 is considered over the spatial domain $\Gamma[-2,4]$, where the initial and boundary conditions are obtained from the exact solution.
 Here, $L$ is the characteristic width of the kernel and assumed to be equal to 0.4 in all test cases.
 For this particular problem, evolution of the profile $u(t,x)$ (from a given Gaussian initial profile) obtained from the proposed invariant scheme (SYM), standard FTCS scheme and compact finite difference (COMP) scheme is depicted in figure \ref{fig3} (left figure). 
 The spatial distribution of errors obtained from these numerical solutions is also shown in this figure (right figure). The invariant compact scheme appears to be capturing the wave propagation significantly better than the FTCS scheme, and slightly better than the compact scheme. Additionally, $L_\infty$ error and root mean square error measures corresponding to the proposed invariant compact scheme, FTCS scheme and standard compact finite difference scheme are presented in table \ref{table2}. It appears that the invariant compact scheme is two orders of magnitude more accurate than the FTCS scheme and is one order of magnitude more accurate than the standard compact finite difference scheme.

\begin{figure}[htbp!] 
  	\centering
  	\includegraphics[width=0.75\textwidth]{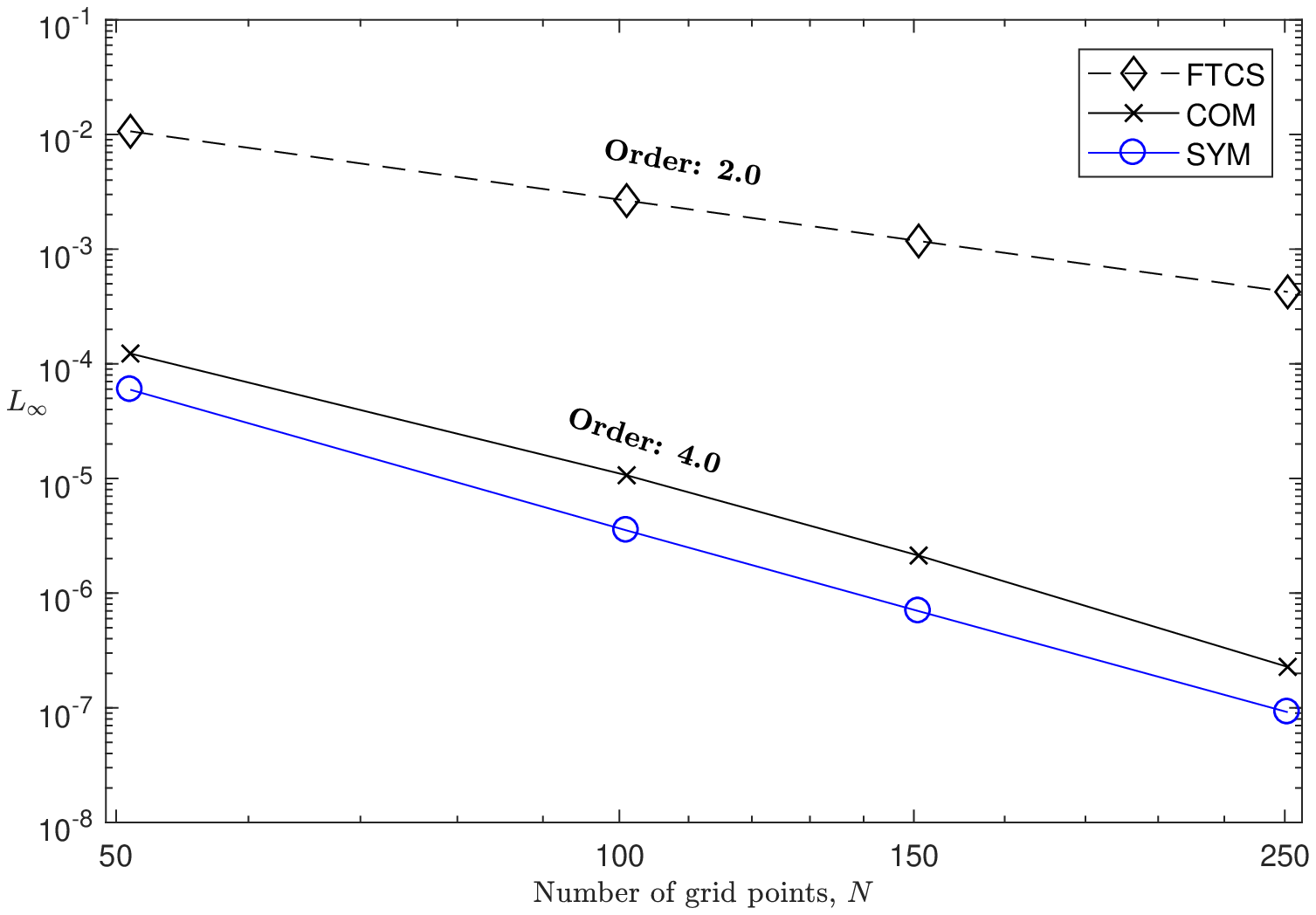} 
  	\captionsetup{format=hang}
  	\caption{Advection-diffusion equation in 1D. Comparison of $L_\infty$ errors of numerical schemes as a function of number of grid points.}
  	\label{fig4}
  \end{figure}

 Additionally, figure \ref{fig4} shows the variation of $L_\infty$ errors associated to the invariant compact scheme, FTCS scheme, and standard noninvariant compact scheme with respect to the number of spatial grid points. The invariant scheme appears to be two orders more accurate than the standard second order FTCS scheme. Moreover, although both the invariant and standard non-invariant compact schemes are fourth order accurate, the invariant scheme appears to have slightly less numerical error.

 In our next test case, we considered the viscous Burgers' equation and developed a fourth order accurate invariant compact scheme that preserve the whole symmetry group, given in Eq.~\eqref{vbesymmetries}, associated with this PDE. The following analytical solution 
 \begin{align} \label{vbeexact}
  u(t,x)=-\frac{2\nu}{\phi}\:\frac{\partial \phi}{\partial x}+4\:,
 \end{align}
 \begin{align*} 
  \phi=\exp \Biggl (-\frac{(x-4t)^2}{4\nu(t+1)}   \Biggr )+\exp \Biggl (-\frac{(x-4t-2\pi)^2}{4\nu(t+1)}   \Biggr )
 \end{align*}
  is considered over the spatial domain $\Gamma[0,2\pi]$ where the initial and boundary conditions are determined from this solution.
  \begin{table}[t]
  	\captionsetup{width=.70\textwidth}
  	\captionsetup{skip=1pt}
  	\centering
  	\caption{Root mean square error (RMSE) and $L_\infty$ error associated with numerical solutions for one-dimensional linear advection-diffusion equation.}\label{table2}
  	\renewcommand{\arraystretch}{1.2}   
  	\small
  	\begin{tabular}{@{\hspace{1.0em}}c@{\hspace{1.5em}}@{\hspace{3.1em}}c@{\hspace{3.1em}}
  			@{\hspace{3.0em}}c@{\hspace{3.0em}}@{\hspace{3.1em}}c@{\hspace{1.1em}}}
  		\hline
  		Error & FTCS & COMP & SYM \\ \hline
  		$L_\infty$ & $2.9 \times 10^{-2}$ & $1.2 \times 10^{-3}$ & $4.6 \times 10^{-4}$ \\ 
  		RMSE & $1.2 \times 10^{-2}$ & $3.7 \times 10^{-4}$ & $2.1 \times 10^{-4}$ \\ \hline
  	\end{tabular}
  \end{table}
  %
  %
  %

\begin{figure}[t] 
  \centering
  \includegraphics[width=0.48\textwidth]{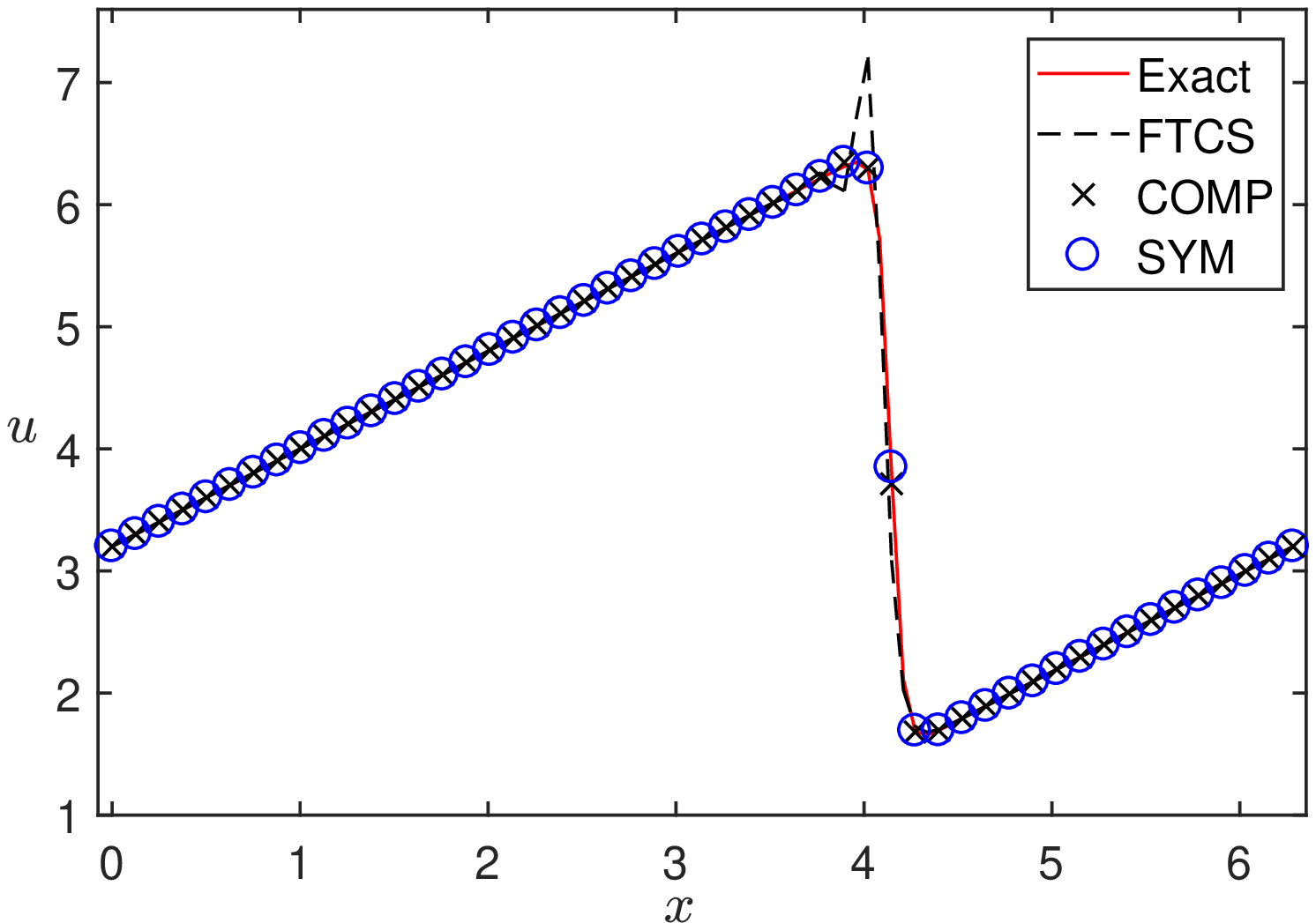} 
  \hfill
  \includegraphics[width=0.48\textwidth]{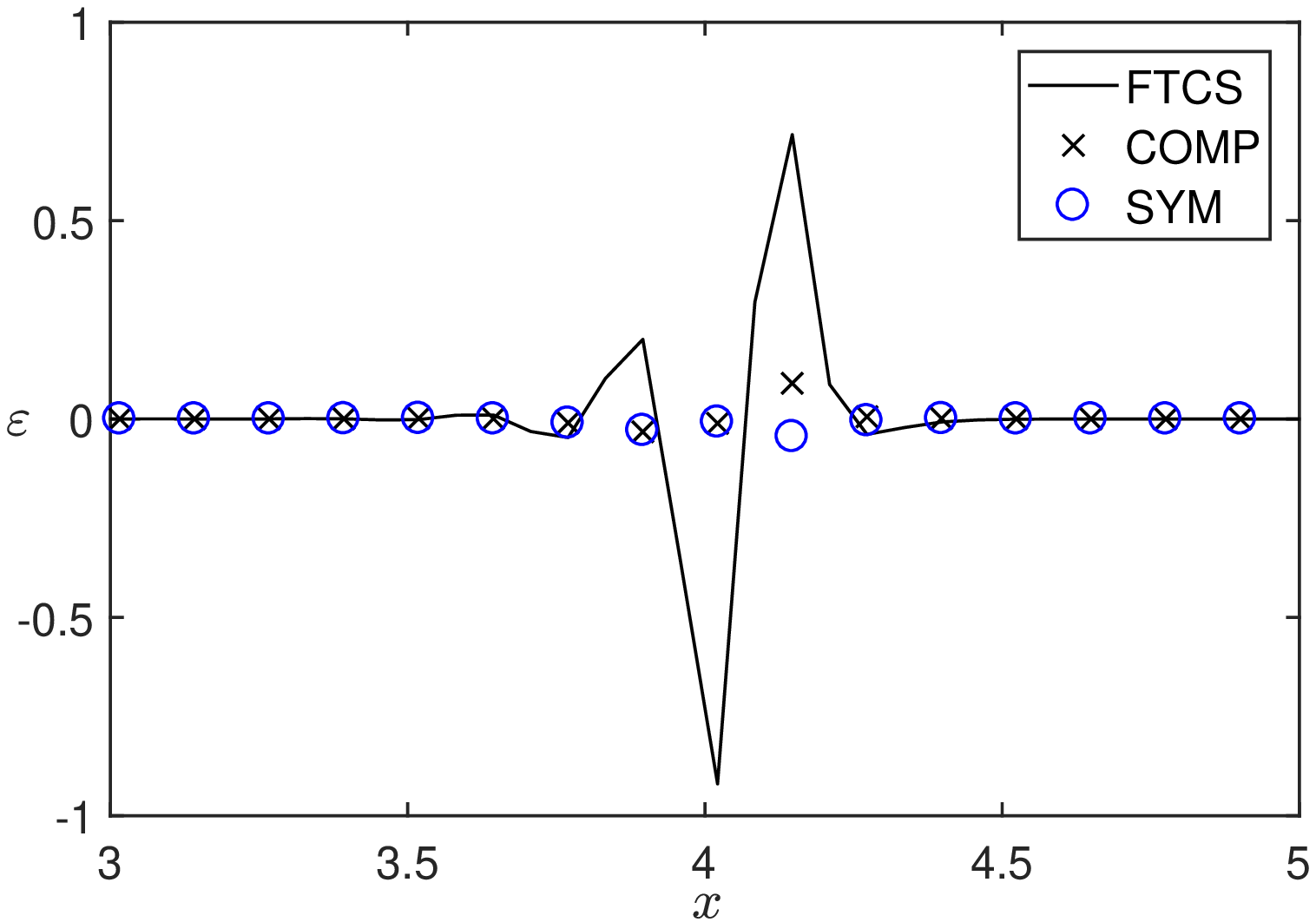} 
  \captionsetup{format=hang}
  \caption{Viscous Burgers' equation. Snapshots of shock formation profiles, at $t=0.25$, obtained from the analytical solution (Exact), the standard forward in time central in space scheme (FTCS), the standard compact scheme (COMP), and the proposed invariant compact scheme (SYM) are shown in the left figure. Spatial distribution of errors for these numerical schemes is displayed in the right figure. Parameter settings:\,\,$h=0.063$,\,\,$\tau=0.0001$,\,\, $\nu=1/12$.~.}
   \label{fig5}
  \end{figure}

  Snapshots of the propagating shock, at $t=0.25$, along with the spatial distribution of numerical errors, obtained from the fourth order accurate invariant compact scheme (SYM), standard second order FTCS scheme, and non-invariant fourth order compact finite difference scheme (COMP) are depicted in figure \ref{fig5}. Although a coarse grid with 101 nodes is used for this particular run, it appears that the invariant scheme performs well and captures the shock propagation significantly better than the standard FTCS scheme, particularly near the shock-front. Further, $L_\infty$ error and root mean square error analysis given in table \ref{table3} also confirms that the invariant compact scheme performs better than the standard FTCS scheme. For this particular run, root mean square errors corresponding to the invariant compact scheme, standard FTCS scheme, and non-invariant compact finite difference scheme are found to be 0.0140, 0.1251, and 0.0143, respectively. Similarly, $L_\infty$ errors of these schemes are determined as 0.1060 (SYM), 0.8962 (FTCS), and 0.0994 (COMP).
 \begin{table}[t]
    \captionsetup{width=.60\textwidth}
   \captionsetup{skip=1pt}
    \centering
    \caption{Root mean square error (RMSE) and $L_\infty$ error associated with numerical solutions for the viscous Burgers' equation.}\label{table3}
    \renewcommand{\arraystretch}{1.2}   
    \small
    \begin{tabular}{@{\hspace{1.0em}}c@{\hspace{1.5em}}@{\hspace{3.1em}}c@{\hspace{3.1em}}
    @{\hspace{3.0em}}c@{\hspace{3.0em}}@{\hspace{3.1em}}c@{\hspace{1.1em}}}
    \hline
    Error & FTCS & COMP & SYM \\ \hline
    $L_\infty$ & $0.8962$ & $0.0994$ & $0.1060$ \\ 
    RMSE & $0.1251$ & $0.0143$ & $0.0140$ \\ \hline
    \end{tabular}
 \end{table}
 %
   %
  %
  %
 %
 %
  \begin{figure}[h] 
  \centering
  \includegraphics[width=0.75\textwidth]{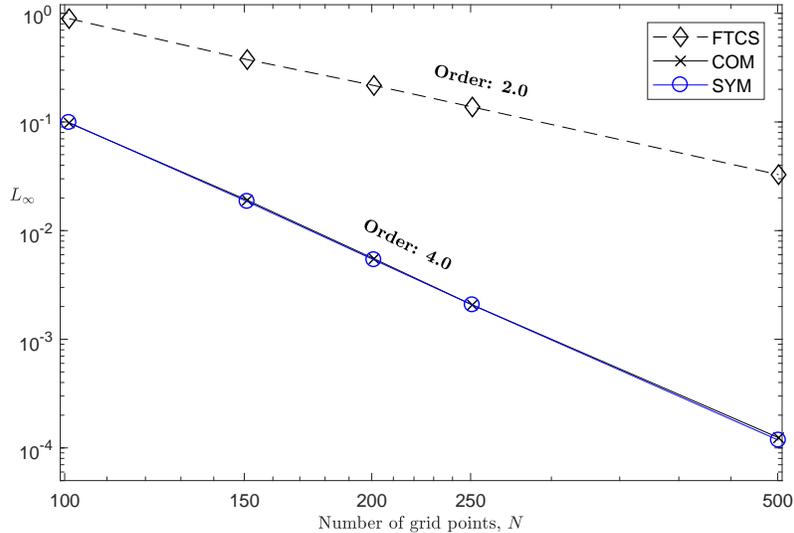} 
  \captionsetup{format=hang}
  \caption{Viscous Burgers' equation. Comparison of $L_\infty$ errors of numerical schemes as a function of number of grid points.}
   \label{fig6}
  \end{figure}

 The variation of $L_\infty$ errors obtained from these numerical schemes with respect to number of spatial grid points is shown in figure \ref{fig6}. As expected, the results obtained from the invariant scheme are indeed fourth order accurate and are two orders more accurate than the standard FTCS scheme, which is known to be a second order accurate scheme. Also, both the invariant scheme and the standard fourth order compact scheme yield results of comparable order of accuracy with negligible differences.

\begin{figure}[t] 
  	\centering
  	\includegraphics[width=0.49\textwidth]{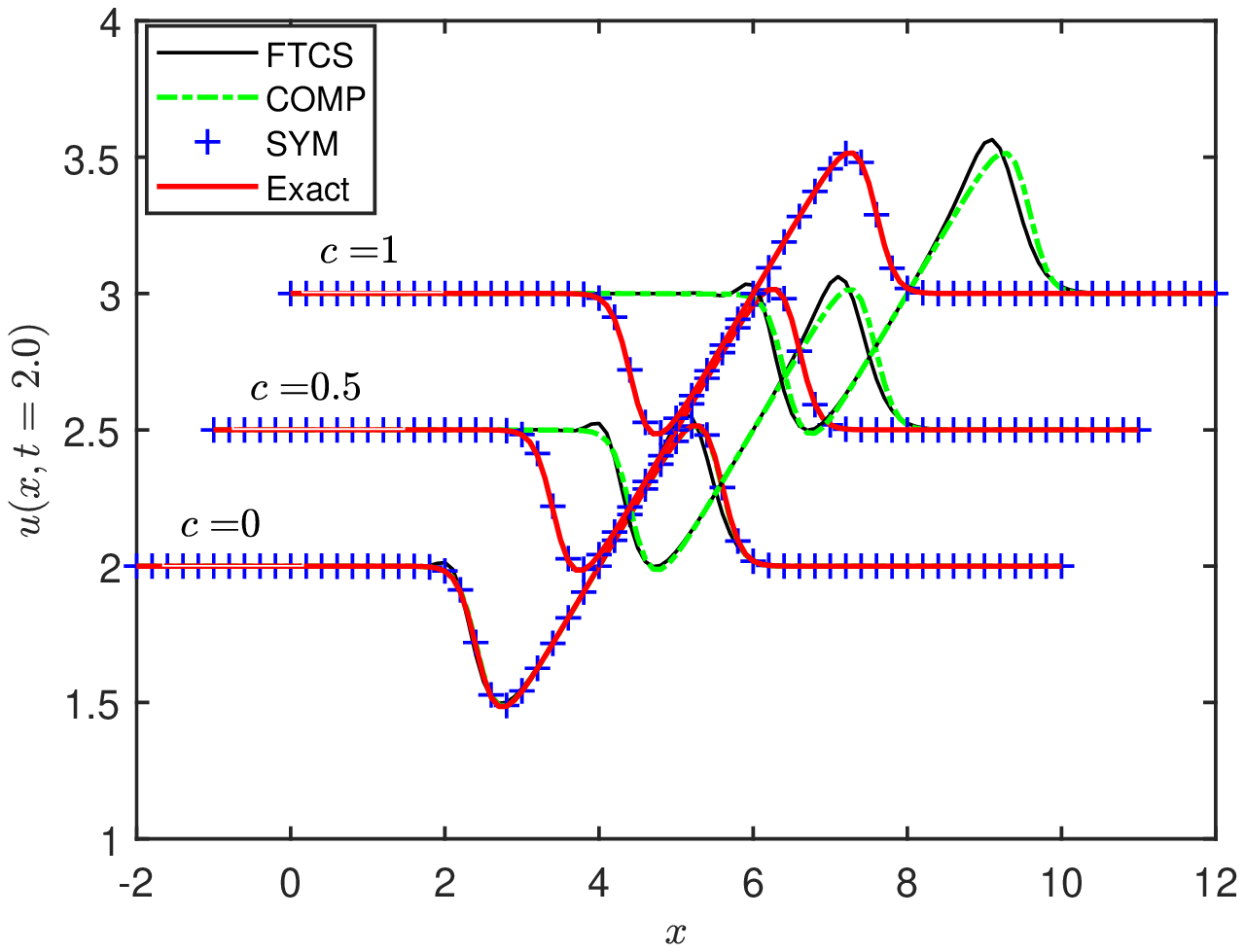} 
  	\hfill
  	\includegraphics[width=0.49\textwidth]{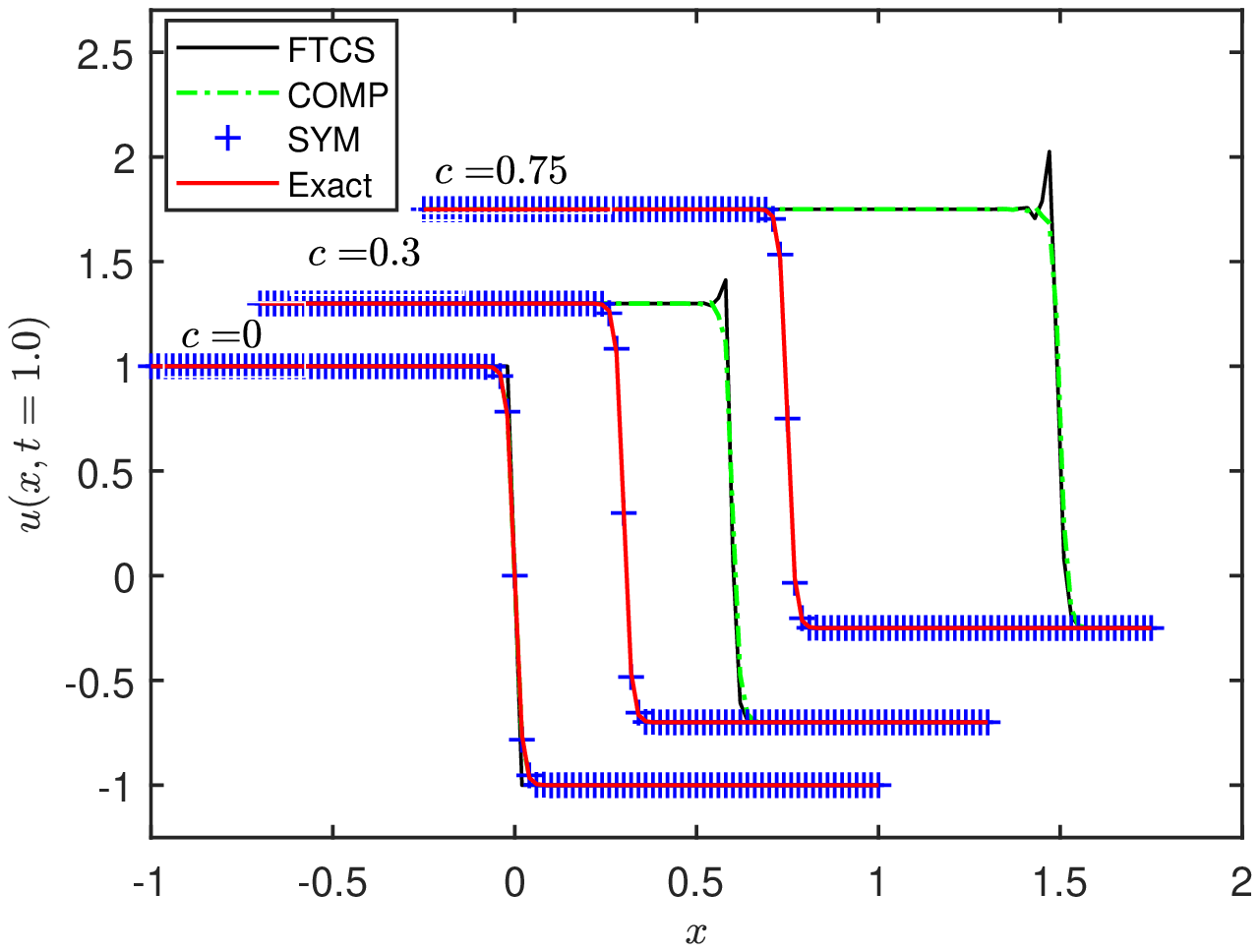} 
  	\captionsetup{format=hang}
  	\caption{Viscous Burgers' equation. Snapshots of numerical solutions, obtained from the analytical solution (Exact), standard forward in time central in space scheme (FTCS), standard compact scheme (COMP), and proposed invariant compact scheme (SYM), evolving from various initial profiles for different values of the Galilean parameter $c$. \textbf{Left:}\,\,$h=0.1$,\,\,$\tau=0.0001$,\,\, $\nu=0.05$,\,\, \textbf{Right:}\,\,$h=0.02$,\,\,$\tau=0.0005$,\,\, $\nu=0.01$.}
  	\label{figVal}
  \end{figure}

  As the proposed invariant compact scheme given in Eq.~\eqref{vbeinvsch} preserves all the symmetry groups of the viscous Burgers' equation (including the Galilean symmetry group), under transformations based on these symmetry groups, the invariant scheme is expected to perform significantly better than the standard numerical schemes that do not preserve these symmetry groups. For instance, under a Galilean transformation of the form
 \begin{align} \label{GalileanTrans}
      \hat x = x + c\:t, \quad \quad \hat t = t, \quad \quad  \hat u = u + c
 \end{align}
  the proposed invariant scheme (SYM) is likely to capture the evolution of the velocity profile significantly better than both the standard FTCS and compact schemes. This is expected as the invariant scheme preserves the Galilean transformation group $X_2$ given in Eq.~\eqref{vbesymmetries} whereas the standard schemes do not. To test this particular advantage of the invariant scheme, we applied the Galilean transformation given in Eq.~\eqref{GalileanTrans} to these numerical schemes and presented the snapshots of the evolution of the numerical solutions from (two different) given initial profiles in figure \ref{figVal}. Additionally, root mean square errors and $L_\infty$ errors associated with these numerical solutions are given in table \ref{tableValleft} and table \ref{tableValright}. These particular initial conditions along with the associated analytical solutions considered for the left and right plots in figure \ref{figVal} can be found in reference \cite{chhay2010}. Based on figure \ref{figVal} and relevant error tables, it appears that when the Galilean parameter $c$ is equal to zero, all the numerical scheme captures the evolution of the solution well which is expected. However, for the cases when the Galilean parameter $c$ is nonzero, both the standard FTCS scheme and compact finite difference scheme appear to overpredict the solution leading to a significant lag in the solution, particularly for large values of $c$. On the other hand, the invariant scheme, as it preserves the Galilean symmetry group, captures the evolution of the solution well even for nonzero values of the Galilean parameter $c$. In fact, in the case of a numerical precision considered in table \ref{tableValleft} and table \ref{tableValright}, the results obtained from the invariant scheme for nonzero values of $c$ are found to be identical to the results of the case where $c=0$. The latter indicates that the Galilean invariance property of the viscous Burgers' equation is indeed preserved in the relevant difference equation. This property of symmetry preservation in numerical schemes can be particularly useful when differential equations associated to more complex symmetries are solved through difference equations.
  \begin{table}[t]
  	\floatsetup{floatrowsep=qquad, captionskip=10pt}
  	\begin{floatrow}[2]
  		\ttabbox%
  		{
  			\renewcommand{\arraystretch}{1.2}    
  			\small
  			\begin{tabularx}{0.45\textwidth}{c *{5}{>{\centering\arraybackslash}X}}
  				\hline
  				$c$ & Error & FTCS & COMP & SYM \\
  				\hline
  				\multirow{2}{*}{$0$}   & $L_\infty$ & $0.1157$ & $0.0100$ & $0.0120$\\ 
  				& RMSE & $0.0213$ & $0.0023$ & $0.0022$\\
  				\hline
  				\multirow{2}{*}{$0.5$}   & $L_\infty$ & $0.5543$ & $0.5131$ & $0.0120$\\ 
  				& RMSE & $0.2424$ & $0.2417$ & $0.0022$\\
  				\hline
  				\multirow{2}{*}{$1.0$}   & $L_\infty$ & $0.9033$ & $0.9166$ & $0.0120$\\ 
  				& RMSE & $0.3232$ & $0.3206$ & $0.0022$\\
  				\hline
  		\end{tabularx}}
  		{\caption{Variation of RMSE and $L_\infty$ errors associated with numerical solutions presented in figure \ref{figVal} (left) with respect to the Galilean parameter $c$ .}
  			\label{tableValleft}}
  		\hfill%
  		\ttabbox%
  		{
  			\renewcommand{\arraystretch}{1.2} 
  			\small
  			\begin{tabularx}{0.45\textwidth}{c *{5}{>{\centering\arraybackslash}X}}
  				\hline
  				$c$ & Error & FTCS & COMP & SYM \\
  				\hline
  				\multirow{2}{*}{$0$}   & $L_\infty$ & $0.2384$ & $0.0269$ & $0.0217$\\ 
  				& RMSE & $0.0339$ & $0.0041$ & $0.0034$\\
  				\hline
  				\multirow{2}{*}{$0.3$}   & $L_\infty$ & $2.1117$ & $2.0058$ & $0.0217$\\ 
  				& RMSE & $0.7521$ & $0.7451$ & $0.0034$\\
  				\hline
  				\multirow{2}{*}{$0.75$}   & $L_\infty$ & $2.2750$ & $2.0118$ & $0.0217$\\ 
  				& RMSE & $1.2066$ & $1.2027$ & $0.0034$\\
  				\hline
  		\end{tabularx}}
  		{\caption{Variation of RMSE and $L_\infty$ errors associated with numerical solutions presented in figure \ref{figVal} (right) with respect to the Galilean parameter $c$ .}
  			\label{tableValright}}
  	\end{floatrow}
  \end{table}%
 As our last test case, we considered the two-dimensional linear advection-diffusion equation and constructed two different fourth order accurate invariant compact finite difference scheme (SYM-1 and SYM-2) for this PDE. The main difference between the constructed invariant schemes are that both are developed via selections of different moving frames, and the details of these selections are given in Section \ref{sec:invnumsch}. The objective is to investigate the effect of these selections on the accuracy of the resulting invariant schemes. The following analytical solution
 \begin{align} \label{ade2dexact}
 	u(t,x,y)=\frac{1}{\sqrt{4\,\pi\,(L^2+\nu\:t)}} \exp \left [{-\frac{(x-\alpha\,t)^2+(y-\beta\,t)^2}{4\,\,(L^2+\nu\:t)}}\right ].
 \end{align}
 is used to evaluate the quality of results obtained from the invariant schemes SYM-1 and SYM-2.
 \begin{table}[t]
 	\captionsetup{width=.87\textwidth}
 	\captionsetup{skip=1pt}
 	\centering
 	\caption{Root mean square error (RMSE) and $L_\infty$ error associated with numerical solutions for two-dimensional linear advection-diffusion equation.}
 	\label{table4}
 	\renewcommand{\arraystretch}{1.2}   
 	\small
 	\begin{tabular}{@{\hspace{1.0em}}c@{\hspace{1.5em}}@{\hspace{3.1em}}c@{\hspace{3.1em}}
 			@{\hspace{3.0em}}c@{\hspace{3.0em}}@{\hspace{3.1em}}c@{\hspace{1.1em}}@{\hspace{3.1em}}c@{\hspace{1.1em}}}
 		\hline
 		Error & FTCS & COMP & SYM-1 & SYM-2 \\ \hline
 		$L_\infty$ & $2.4 \times 10^{-3}$ & $3.8 \times 10^{-5}$ & $3.4 \times 10^{-5}$ & $3.3 \times 10^{-5}$ \\ 
 		RMSE & $2.7 \times 10^{-4}$ & $3.4 \times 10^{-6}$ & $3.3 \times 10^{-6}$ & $3.1 \times 10^{-6}$ \\ \hline
 	\end{tabular}
 \end{table}
 \begin{figure}[t] 
 	\centering
 	\includegraphics[width=1.0\textwidth]{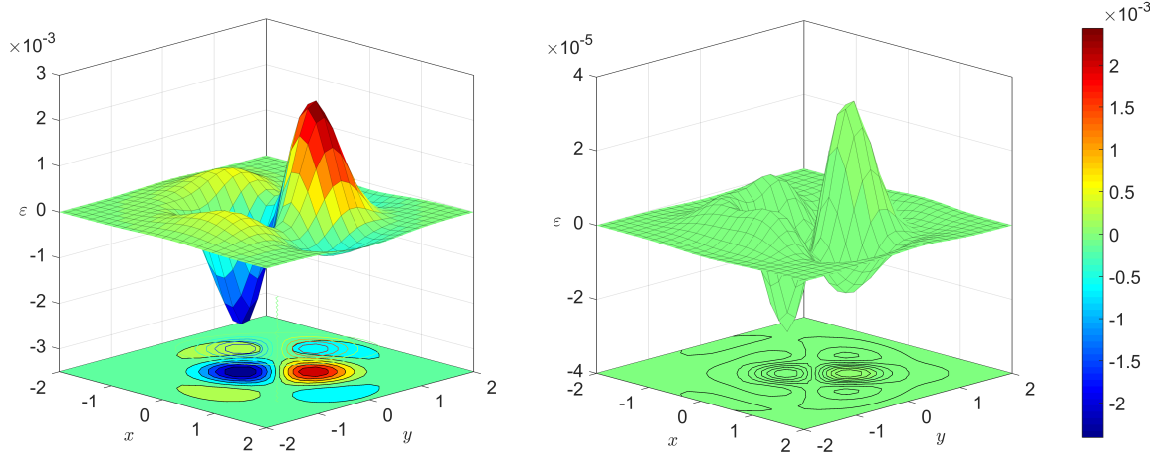}
 	\captionsetup{format=hang}
 	\caption{Linear advection-diffusion equation in 2D. Spatial distribution of numerical errors, at $t=0.1$, obtained from the classical base scheme (left) and the proposed invariant scheme (right). Parameter settings:\,\,$h_x=0.16$,\,\,\,$h_y=0.16$,\,\,\,$\tau=0.0001$,\,\,\, $\alpha=1.0$,\,\,\,$\beta=1.0$,\,\,\,$\nu=1/60$.}
 	\label{fig7}
 \end{figure}
 %
 %
 %
 \begin{figure}[!htbp] 
 	\centering
 	\includegraphics[width=0.75\textwidth]{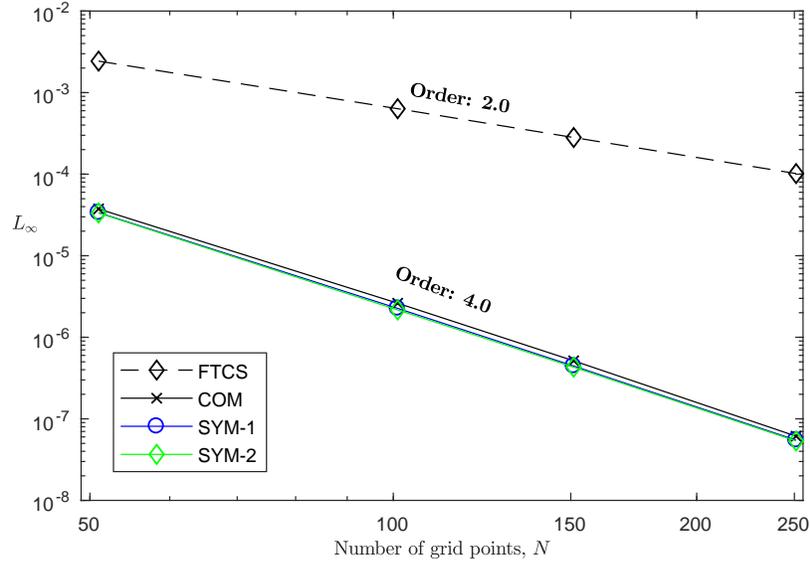} 
 	\captionsetup{format=hang}
 	\caption{Advection-diffusion equation in 2D. Comparison of $L_\infty$ errors of numerical schemes as a function of number of grid points.}
 	\label{fig8}
 \end{figure}

 Spatial distribution of numerical errors corresponding to the proposed invariant compact finite difference scheme (SYM-2) and standard non-invariant FTCS scheme is given in figure \ref{fig7}. 
 Based on this figure, it appears that the invariant scheme has significantly less numerical error compared to the standard non-invariant FTCS scheme in this case as well. 
 This improvement in numerical accuracy is also verified by the error analysis given in table \ref{table4}, where both invariant schemes (SYM-1 and SYM-2) perform better than the standard schemes. 
 $L_\infty$ errors obtained from the invariant scheme SYM-1, invariant scheme SYM-2, FTCS scheme, and standard non-invariant compact scheme are noted as $3.4 \times 10^{-5}$, $3.3 \times 10^{-5}$, $2.4 \times 10^{-3}$, and $3.8 \times 10^{-5}$, respectively. 
 It appears that the invariant schemes are at least two orders of magnitude more accurate than the standard FTCS scheme. 
 Root mean square error measures of these numerical schemes also yield similar results, which are $3.3 \times 10^{-6}$ and $3.1 \times 10^{-6}$  for the invariant schemes SYM-1 and SYM-2, $2.7 \times 10^{-4}$ for the FTCS scheme, and $3.4 \times 10^{-6}$ for the non-invariant compact finite difference scheme.

 The variation of $L_\infty$ errors (obtained from the proposed invariant schemes, standard FTCS scheme, and non-invariant compact scheme) with respect to the number of spatial grid points is presented in figure \ref{fig8}. 
 As expected, both of the proposed invariant compact schemes constructed for the two-dimensional linear advection-diffusion equation are indeed fourth order accurate, and perform significantly better than the second order standard forward in time central in space finite difference scheme (FTCS). 
 Moreover, these invariant schemes also perform with slightly less error compared to the non-invariant compact scheme which is known to be a fourth order accurate scheme. 
 Further, the invariant scheme SYM-2 appears to be slightly more accurate than the invariant scheme SYM-1 which indicates that the selection of moving frames could affect the accuracy of resulting invariant schemes.
 Although for this particular problem, the differences in the results obtained from the invariant schemes appear to be minor, in general the moving frames must be chosen carefully.

\section{Conclusion}\label{sec:conc}
Compact finite difference schemes are preferred over standard finite difference schemes as these schemes enable high order accuracy on stencils with comparably small number of grid points, and have good, spectral-like resolution. 
In this paper, we presented a method, that is based on moving frames, for construction of invariant compact finite difference schemes that preserve Lie symmetry groups of underlying partial differential equations.
In this method, we first determine the extended symmetry groups of PDEs, and then obtain point transformations based on these symmetry groups. 
These transformations are then applied to some (non-invariant) base compact finite difference schemes such that all the system variables (i.e., independent and dependent variables) and derivatives of these compact schemes are transformed.
We then determine the unknown symmetry parameters that exist in these symmetry-based point transformations by considering convenient moving frames that are obtained through Cartan's method of normalization.
In most cases, such convenient moving frames not only result in significant improvement in numerical accuracy but also notably simplify the numerical representations of the resulting invariant schemes, and eventually make them easier to program.
Performance of the proposed method was evaluated via construction of high order accurate invariant compact finite difference schemes (built on simple three-point stencils) for some linear and nonlinear PDEs. Based on our evaluations, we concluded that symmetry preservation has the potential to significantly improve numerical accuracy of compact schemes, besides embedding important geometric properties of underlying PDEs.

As our first test case, we considered the inviscid Burgers' equation and constructed a high order accurate invariant compact finite difference scheme for this PDE. 
Although the order of accuracy of compact schemes can be arbitrarily set by considering suitable compact finite difference algorithms, for this particular problem, we chose fourth order accurate compact algorithms to approximate the spatial derivatives and constructed an invariant scheme based on these algorithms.
In all the test problems, the temporal derivatives were handled through standard forward differencing.
For this particular PDE, in order to improve the numerical accuracy from first to second order in time, the base scheme was modified using defect correction techniques. 
The results obtained from this fourth order accurate invariant compact scheme were found to be slightly better than the results obtained from the standard compact scheme and were notably better than those of the standard FTCS scheme. 
%
%
For all the test cases, the computation times, required to run a simulation with a numerical error of comparable order, were found to be similar for both the proposed invariant scheme and standard compact scheme, and the differences were negligible.

As our next test problem, we considered the one-dimensional linear advection-diffusion equation and developed a fourth order accurate invariant compact scheme for this problem as well.
For this particular problem, through the use of convenient moving frames (i.e., $\tilde u_{\tilde x \tilde x}=0$), the numerical representation of the base scheme were reduced to a form of the linear advection equation ($\tilde u_{\tilde t} + \alpha \tilde u_{\tilde x} =0$) in the transformed space. 
Similar to the previous problem, the quality of results obtained from this invariant compact scheme (in terms of numerical accuracy) was found to be better than that of the standard FTCS and compact schemes. 

Next we constructed a fourth order accurate invariant compact finite difference scheme for the viscous Burgers' equation (which is of the form of a linear heat equation, $\tilde u_{\tilde t}=\nu \tilde u_{\tilde x \tilde x}$, in the transformed space for the normalization condition $\tilde u_{\tilde x}=0$) that preserves all the symmetries of the Burgers' equation, and compared our results with the standard schemes. 
As expected, the proposed invariant compact scheme developed for this problem yielded more accurate results than standard schemes in this case as well. 
In particular, the performance of the proposed invariant scheme was significantly better than that of the standard schemes when a Galilean transformation is applied to these schemes (see figure ~\ref{figVal} and tables \ref{tableValleft}--\ref{tableValright}) to test how these schemes are affected by such transformations that are based on symmetries of the underlying differential equation. 
This is due to the fact that the invariant scheme preserves the Galilean symmetry group of the viscous Burgers' equation, whereas the standard schemes do not.

In order to demonstrate the implementation of the proposed method to a multidimensional problem, as our last test case, we considered the two-dimensional linear advection-diffusion equation and constructed a couple of fourth order accurate invariant compact schemes for this problem, where different moving frames are used in the construction of each invariant scheme to evaluate how this action effects the accuracy of the resulting schemes.
For the first invariant scheme SYM-1, a normalization condition of the form $\tilde u_{\tilde x \tilde x} = 0$ is used to determine the projection group parameter $s_1$ whereas for the other invariant scheme (SYM-2), this particular parameter was determined using the normalization condition $\tilde u_{\tilde x \tilde x} +\tilde u_{\tilde y \tilde y} = 0$. 
Although both normalization conditions simplify the base compact scheme considered for this PDE notably, the latter condition reduces the base scheme to the form of a two-dimensional linear advection equation ($\tilde u_{\tilde t} + \alpha \tilde u_{\tilde x}+ \beta \tilde u_{\tilde y}=0$) in the transformed space.
As for the results obtained from these invariant schemes, SYM-2 appears to be slightly more accurate than SYM-1 where both of these schemes are notably more accurate than standard schemes. 
Although for this particular problem, selection of different moving frames in the construction of invariant schemes did not affect the accuracy of these schemes significantly, this may not be the case for other problems as there are usually infinitely many applicable moving frames, and not all of them will result in accurate invariant schemes. 

While the proposed method could be effectively used for construction of invariant compact finite difference schemes with desired order of accuracy, there are few issues that need to be addressed in more detail. 
Further research is required to understand how the performance of invariant compact schemes (constructed through the proposed method) is affected by the choice of subgroups (considered for preservation in the difference equation), choice of moving frames among infinite number of possibilities, and the nature of initial/boundary conditions and their compatibility with the selected subgroups. 
Based on our simulations, we observed that although it is possible to consider the whole symmetry group of a PDE for preservation in difference equations, this often leads to cumbersome numerical representations without notably enhancing numerical accuracy. 
For instance, in the case of the viscous Burgers' equation, the whole symmetry group of the PDE is preserved in the related difference equation. However, the advantages owing to the inclusion of the Galilean subgroup only become significant when the invariant scheme is actually transformed under a Galilean transformation as demonstrated in figure \ref{figVal}.
Further, the choice of moving frames which are used to determine the unknown group parameters could affect the accuracy of resulting invariant schemes. 
To our knowledge, there is no systematic approach to select the best moving frame and one must consider all the pros and cons of a particular moving frame before making a selection.
Based on our observations, we found that a moving frame that removes the leading order terms from truncation error of a difference equation is usually a good choice as such a moving frame also simplifies the base scheme (in the transformed space) and make it easier to program.
Moreover, the performance of the constructed invariant schemes might be affected by the chosen initial/boundary conditions, especially if these conditions are not compatible with the chosen subgroups.
This might be due to the fact that some of the limitations of base difference equations carry over to the constructed invariant schemes.
For instance, for cases where discontinuities develop in solutions, the performance of the constructed invariant schemes will undoubtedly depend on the chosen base numerical schemes.
This obstacle could be avoided by selecting base schemes that are better suited to handle such discontinuities.

\section*{Acknowledgments}
The first author (PhD candidate) is grateful for financial support provided by the Ministry of National Education of Turkey.


\bibliographystyle{elsarticle-num}
\bibliography{mainArchive}

\end{document}